\documentclass[%
 reprint,
 amsmath,amssymb,
 aps,
]{revtex4-2}

\usepackage{graphicx}
\usepackage{dcolumn}
\usepackage{bm}
\usepackage{subcaption}
\usepackage{makecell}
\usepackage{multirow}

\begin{document}

\preprint{APS/123-QED}

\title{The magnetic properties of the iron phthalocyanine molecule grafted to the Ti\textsubscript{2}C MXene layer}

\author{Aleksei Koshevarnikov}
 \email{aleksei.koshevarnikov@fuw.edu.pl}
 \author{Tomi Ketolainen}
\author{Jacek A. Majewski}%
 \email{jacek.majewski@fuw.edu.pl}

\affiliation{%
Institute of Theoretical Physics, Faculty of Physics,
University of Warsaw, Pasteura 5, 02-093 Warsaw, Poland
}%


\begin{abstract}

The magnetic tetrapyrrole molecules (such as porphyrins and phthalocyanines) with an active transition metal atom in their centre are currently intensively studied as prosperous potential elements of devices for high-density information storage and processing. It has been recently proved that by means of external factors one could induce two stable fully controllable molecular states. Therefore, hybrid systems consisting of such magnetic molecules and suitable carriers from the family of two-dimensional materials are often considered as promising highly scalable spintronic systems that could in the near future lead to novel industrial applications. Here, we perform the spin polarised density functional theory (DFT) studies of the hybrid system, which is the iron phthalocyanine molecule (FePc) on the top of the titanium carbide Ti\textsubscript{2}C MXene layer. The most relevant issue in this part is the interaction between magnetic atoms: Ti from MXene substrate and iron from FePc. Four various magnetic configurations of FePc/Ti\textsubscript{2}C were considered. The significant ferromagnetic interaction between the iron atom and the upper titanium layer plays important role in the reorientation of the iron atom's magnetic moment. We also analyse a model of the system in which the FePc molecule is in a quintet state (the ground state of an isolated molecule is a triplet). To get a better understanding of the physics of the FePc/Ti\textsubscript{2}C hybrid system, we studied the hybrid systems with a single iron atom and non-magnetic H\textsubscript{2}Pc on the Ti\textsubscript{2}C layer, Fe/Ti\textsubscript{2}C and H\textsubscript{2}Pc/Ti\textsubscript{2}C, respectively, which nicely explains the role of the Pc ligand in the FePc/Ti\textsubscript{2}C hybrid system. 

\end{abstract}

\maketitle


\section{Introduction}

\begin{figure}
\centering
\includegraphics[width=\columnwidth]{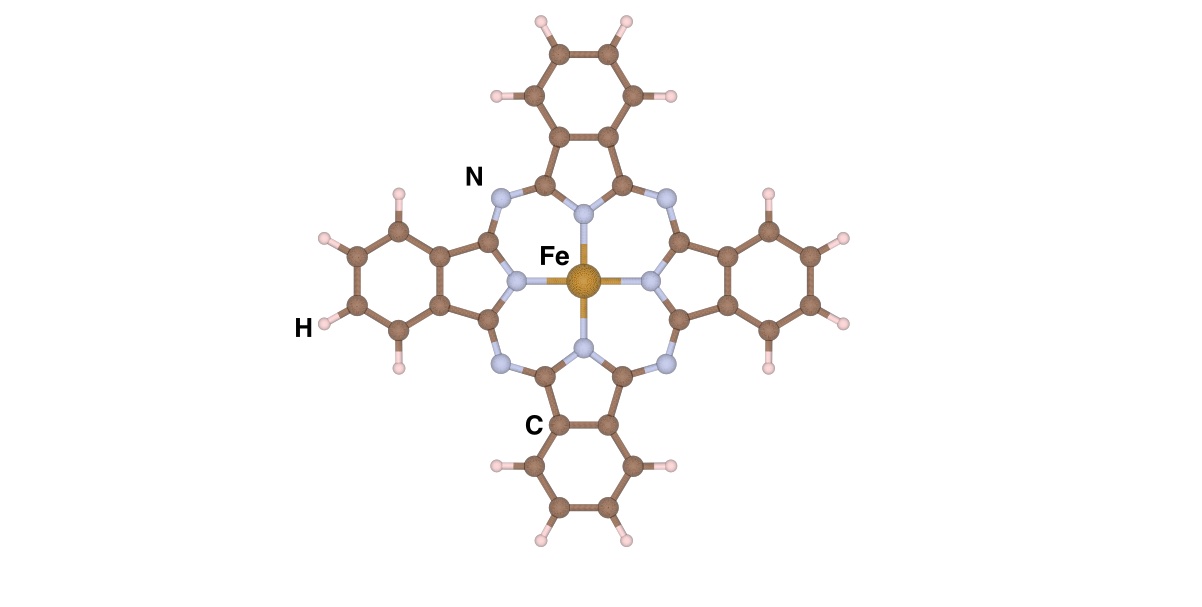}
\caption{The FePc molecule}
\label{fig:FePc}
\end{figure}

Transition metal phthalocyanines (TMPc’s) (Fig. \ref{fig:FePc}) are widely known organometallic compounds that have been well known since the 1930s.\cite{linstead1934212} They consist of isoindole rings interconnected via an $sp^2$-hybridized nitrogen atom and a transition metal atom in the centre. The first commercial implementation of these compounds was as a blue pigment. Over the years, this type of molecule has attracted attention of researches researchers in the field of molecular electronics. The ease of preparation, the simple molecular structure, and the possibility of functionalisation have made TMPc’s one of the leading species for applications in the modern fields of spintronics and optoelectronics.\cite{lian2010printed,barraud2016phthalocyanine,warner2013potential} Such molecules can also act as a part of an organic photodetector,\cite{kaneko2003fast} and as a hole injection layer in OLED displays.\cite{tadayyon2004cupc}

\begin{figure}
\centering
\begin{subfigure}{0.75\linewidth}
        \centering
    \includegraphics[width=\linewidth]{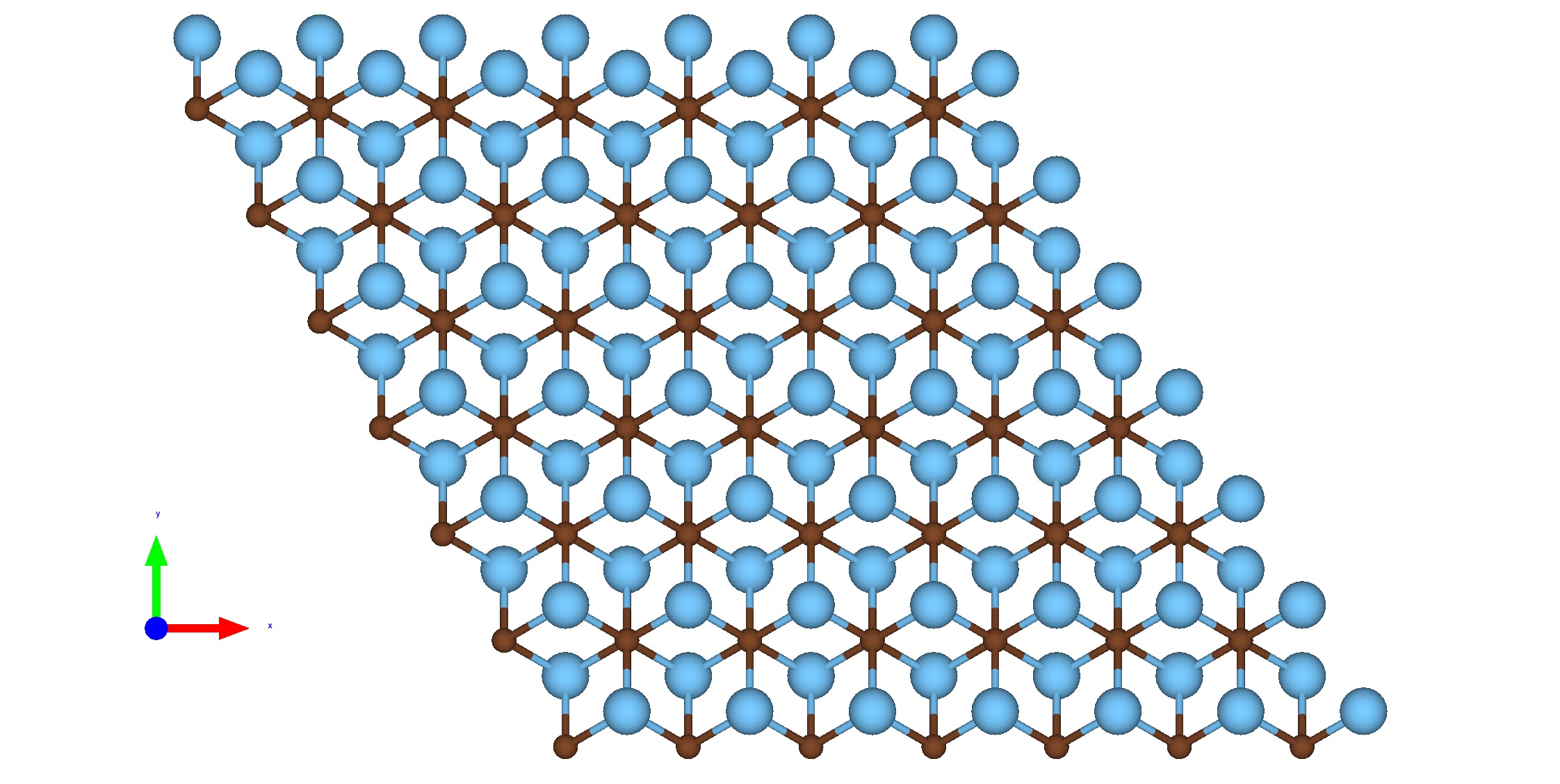}
    \caption{}
    \label{fig:Ti2CBig}
\end{subfigure}
\begin{subfigure}{0.49\linewidth}
        \centering
    \includegraphics[width=\linewidth]{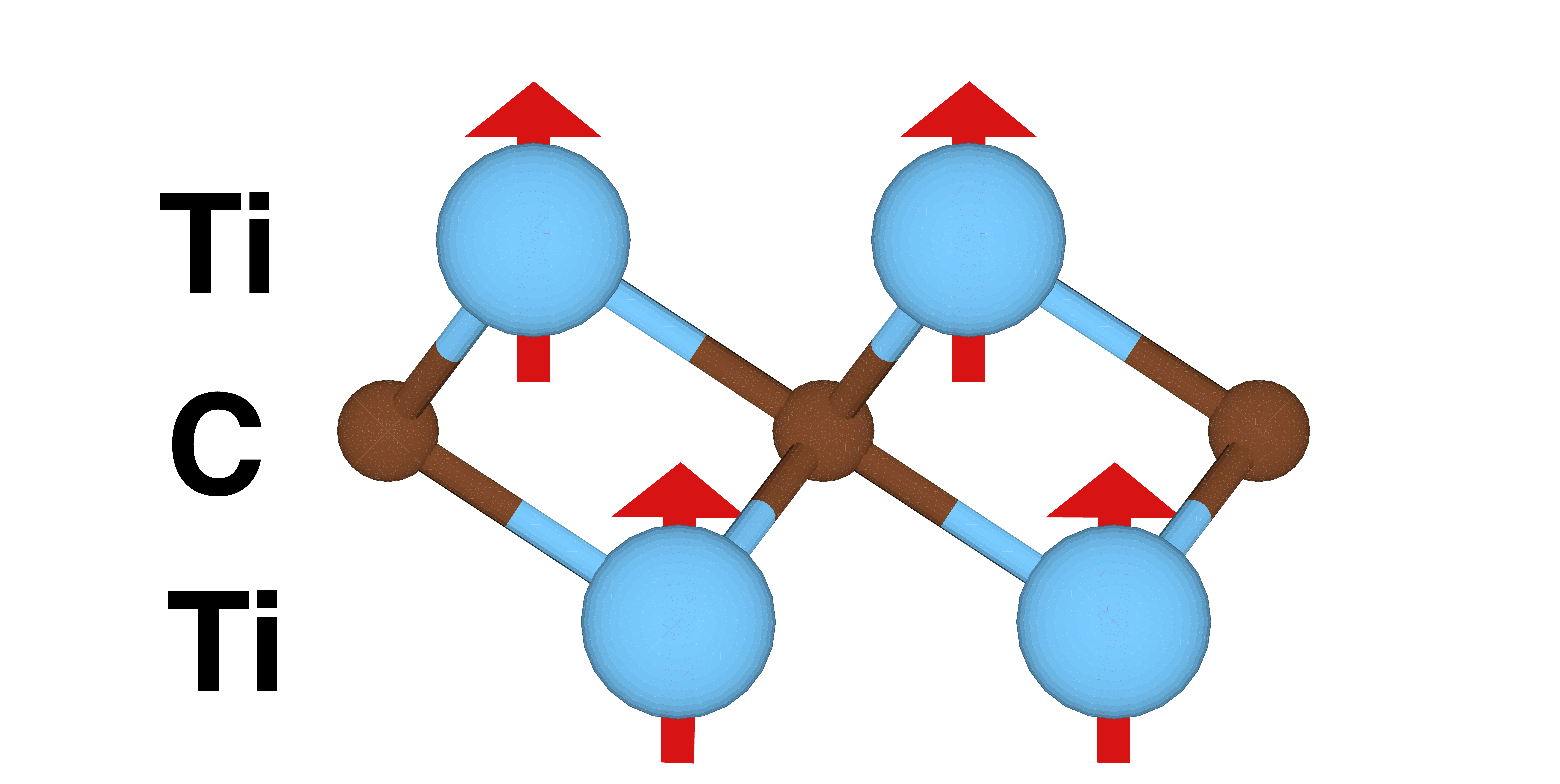}
    \caption{}
    \label{fig:Ti2CFM}
\end{subfigure}
\begin{subfigure}{0.49\linewidth}
        \centering
    \includegraphics[width=\linewidth]{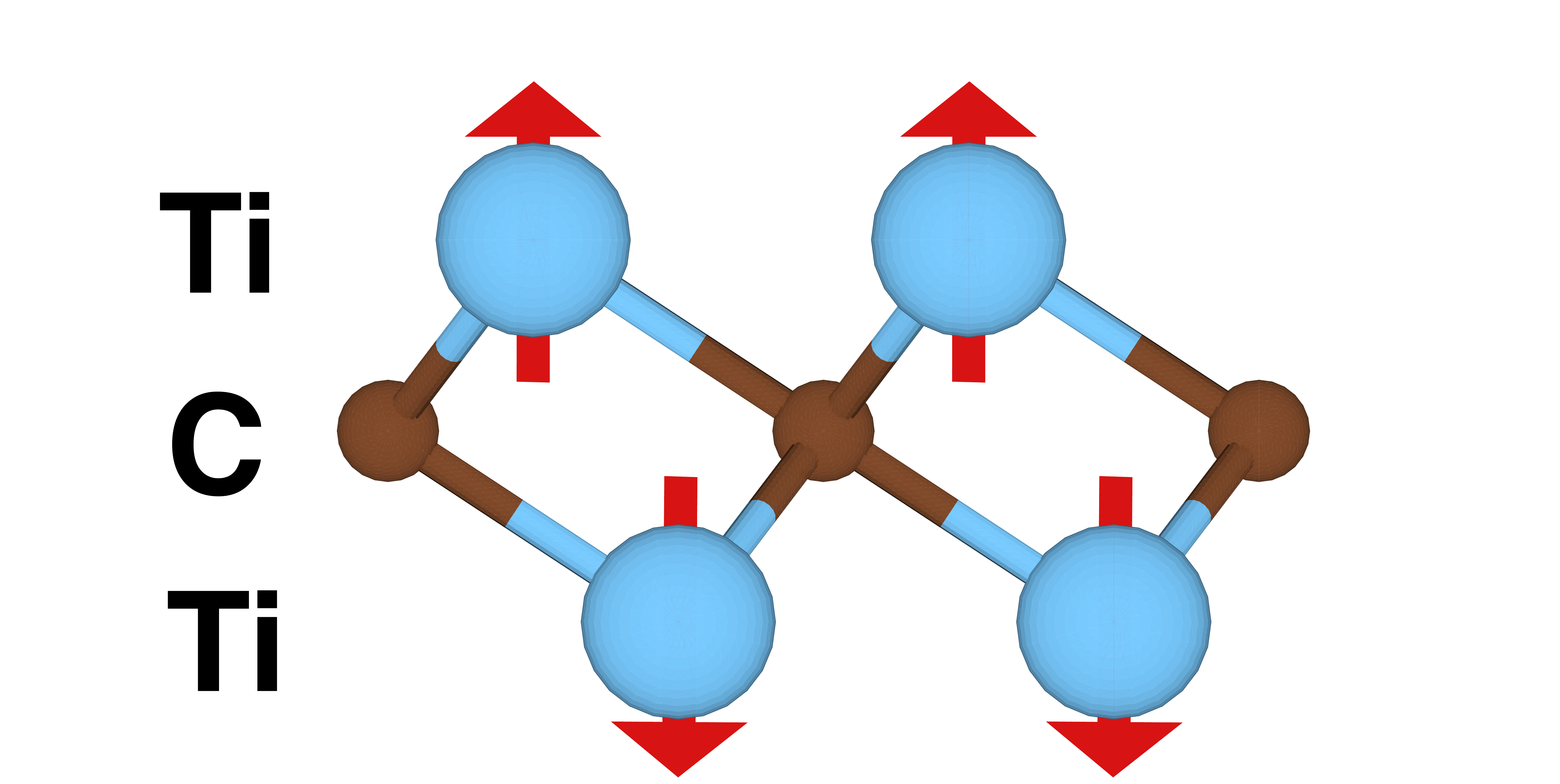}
    \caption{}
    \label{fig:Ti2CAFM}
\end{subfigure}
    \caption{The Ti\textsubscript{2}C 2D-threelayer system: (a) view from above; and side views of (b) ferromagnetic and (c) antiferromagnetic configurations. Red arrows indicate the internal magnetic moment of Ti atoms. }
    \label{fig:Ti2C}
\end{figure}

MXene is a class of 2D compounds obtained from MAX phases (layered hexagonal carbides or nitrides with a formula M\textsubscript{n+1}AX\textsubscript{n}, where n=1-3, M indicates a transitional metal, A - elements mostly from groups 13 and 14, and X - C or N atoms) by etching A elements. Firstly discovered experimentally in 2011,\cite{naguib2011two} this class of compounds has been developed rapidly. A lot of new species and their functionalised derivatives have been synthesised. Soon potential applications in thermoelectricity, catalysis, and energy storage have been found.\cite{pang2019applications,fu2019rational} Here, we are interested in the magnetic properties of MXenes and especially of the Ti\textsubscript{2}C (Fig. \ref{fig:Ti2C}). It was shown \cite{lv2020monolayer} that this layer exhibits several magnetic configurations where the most stable are the configurations with co-directional (ferromagnetic, Fig. \ref{fig:Ti2CFM}) and counter-directional (antiferromagnetic, \ref{fig:Ti2CAFM}) magnetic moments of the Ti sublayers. The energy of the antiferromagnetic configuration is predicted to be lower just by 10 meV per primitive cell.

Because MXenes are relatively new materials, there are not many quantum-chemical studies of the adsorption of molecules on the MXene surface. Experimental studies of such systems are hampered by the fact that the outer layers of the material adsorb substances from the atmosphere, changing their chemical structure. Experimental studies of TMPc/MXene hybrid systems were carried out taking into account that the surface is saturated by atmospheric functional groups. In these cases, MXenes lose their magnetic properties. Nonetheless, the FePc/Ti\textsubscript{3}C\textsubscript{2}X\textsubscript{2} system was found to be a good catalyst for redox reactions.\cite{li2018marriage}
In addition, the same complex works well in determining miRNAs and diagnosis of cancer biomarkers.\cite{duan2020construction}
Computational studies of MXenes have gone further than experimental ones. Quantum-chemical methods make it possible to investigate compounds that are still difficult to obtain experimentally.

Theoretical studies of organometallic molecules on MXene surfaces were not found in the literature, but there are several theoretical studies devoted to the adsorption of atoms on the surface of Ti\textsubscript{2}C and Ti\textsubscript{3}C\textsubscript{2} 2D MXene layers. It was found \cite{gao2019functionalization} that the adsorption energies of 3d, 4d, and 5d transition metal atoms on Ti\textsubscript{3}C\textsubscript{2} are in the range of -7.98 to -1.05 eV.
3d-transition metals on M\textsubscript{2}C layers (M = Ti, V, Cr, Zr, Nb, Mo, Hf, Ta, and W) were studied \cite{oschinski2021interaction} as single-atom catalysts to find an alternative to Pt-based catalysts. Also, there are studies about metal atoms on functionalised layers for single-atom catalysts \cite{huang2019single, zhao2019mxene} and Li-ion storage.\cite{wan2018first} It should be noted that these studies did not take into account the spin polarisation of surfaces.
The results of the studies are presented in Table \ref{table:FeTi2CLiterature}. The calculation parameters given in parentheses will be explained in the next chapter. 

\begin{table*}
    \caption{Energetic, geometric and magnetic characteristics of Fe/Ti\textsubscript{2}C and Fe/Ti\textsubscript{3}C\textsubscript{2} hybrid systems. The sources of the data are indicated. Adsorption sites for Ti\textsubscript{3}C\textsubscript{2}: Hcp - on the higher C-atom, Fcc - on the middle Ti; for Ti\textsubscript{2}C: Hcp and Fcc - on the bottom Ti, but with different symmetry surrounding.}
    \label{table:FeTi2CLiterature}
\begin{ruledtabular}
\begin{tabular}{cccc}
  & Adsorption Energy, eV & Fe height from the layer, \AA & $\mu$(Fe)
 \\
\colrule
 Fe on Ti\textsubscript{3}C\textsubscript{2} (PBE)\cite{gao2019functionalization} & \makecell{Hcp: -3.849
 \\
Fcc: -3.847} & \makecell{Hcp: 2.373
\\
Fcc: 2.360} & \makecell{---
\\
---} \\

 Fe on Ti\textsubscript{2}C (PBE+D3)\cite{oschinski2021interaction} & \makecell{Hcp: -4.093
 \\
Fcc: -3.897} & \makecell{Hcp: 1.708
\\
Fcc: 1.644} & \makecell{Hcp: -2.03
\\
Fcc: -2.28} \\
\end{tabular}
\end{ruledtabular}
        \end{table*}

Determination of the magnetic moment of each atom in a system is possible using spin-polarized scanning tunnelling microscopy.\cite{Brede2016,khajetoorians2010detecting,chilian2011experimental} In these works, the Fe atom was studied in the interaction with InSb(110) surface. In this study, the iron atom falls into the surface due to the large lattice constant. In the case of the iron atom on top of the Cu(001) surface, it was shown \cite{schuler2017functionalizing} that the electronic and magnetic properties of adatoms are strongly affected by the tip-surface distance. 

In the purely theoretical study of iron chains on Cu(001) and Cu(111) surfaces \cite{lazarovits2003magnetic} a single iron atom acts as a donor of spin momentum. It retains most of the spin momentum when it is in the so-called "ontop" position and loses it when the atom is in or inside the surface. The Bader charge transfer analysis for 3d TM atoms on graphene and graphene/Ni(111) surfaces \cite{ellinger2021magnetic} indicates that for all elements in the series electrons are transferred from the adatom to the graphene layer, leaving the net charge on the adatom positive.
Magnetic atoms on the surface were also studied. Cobalt and iron atoms were placed onto the Pt(111) surface \cite{blonski2009density} and also on Pt(111) and Ir(111) surfaces.\cite{etz2008magnetic} There was shown that adatoms induce polarisation on nearby surface atoms. The magnetic anisotropy parameters also were calculated and they are in agreement with the experimentally determined ones, where inelastic tunnelling spectroscopy was implemented.\cite{balashov2009magnetic}

The ability of an atom on a surface to have two stable states was studied in several articles. TM atoms on the graphene/Ni(111) surface can be ferromagnetic and antiferromagnetic  toward the surface magnetisation.\cite{ellinger2021magnetic} It was found that for Ti, V and Cr antiferromagnetic alignment is preferable while for Mn, Fe and Co it is ferromagnetic (exchange energy for Fe, in this case, is 10 meV).
By utilizing a combination of scanning tunnelling spectroscopy and DFT methods, it was shown that a Co atom on semiconducting black phosphorus \cite{kiraly2018orbitally} has two states: low-spin and high-spin. It was shown experimentally that the state of the atom can be switched electrically. 
A holmium atom on the MgO surface exhibits bistability property with up and down spin states.\cite{natterer2017reading} It was shown that it is possible to read the states using a tunnel magnetoresistance and induce particular magnetic state of the iron atom (just to "write" it) with current pulses using a scanning tunnelling microscope.

In this paper, we present the study of the FePc molecule on the Ti\textsubscript{2}C MXene layer. Here, we mostly focus on the magnetic properties of the formed system. Different magnetic configurations of the FePc/Ti\textsubscript{2}C hybrid system are modelled and compared. Additionally, a single Fe atom and the H\textsubscript{2}Pc molecule on the top of  Ti\textsubscript{2}C are considered. It allows one to understand the role of the phthalocyanine ligand in the complex. 

\section{Computational Details}

The DFT calculations for the studied hybrid FePc/Ti\textsubscript{2}C system with periodic boundary conditions were performed employing the Quantum Espresso 6.5 numerical package \cite{giannozzi2009quantum} with the generalised gradient approximation (GGA) \cite{langreth1983beyond} realised through the Perdew-Burke-Ernzerhof (PBE) exchange-correlation functional.\cite{perdew1996generalized} The van der Waals interaction between the molecule and the Ti\textsubscript{2}C layer was accounted for within the Grimme DFT-D3 ad-hoc scheme.\cite{grimme2010consistent} Rappe-Rabe-Kaxiras-Joannopoulos (RRKJ) ultrasoft pseudopotentials from pslibrary \cite{dal2014pseudopotentials} were implemented. To treat the strong on-site Coulomb interaction of TM d-electrons, we used the DFT+U approach within the Hubbard model.\cite{anisimov1997first} The U parameter value for the Fe atom (U = 4 eV) was taken from the previous results of linear response calculations for TMPc molecules.\cite{brumboiu2019ligand} To simulate the excited state with S = 2, the Hubbard parameter U was changed from 4 to 6 eV.  The kinetic energy cutoff for wavefunctions was set to 45 Ry and the corresponding parameter for charge density and potential to 650 Ry. Preliminary estimations, optimisation, and calculation of energy parameters were performed at the $\Gamma$ point. The chosen supercell for the FePc/Ti\textsubscript{2}C hybrid system consisted of 7x7 Ti\textsubscript{2}C primitive cells. Such size of the supercell makes it possible to place FePc molecules at a distance of 6.36 \AA\hspace{0pt} from each other, which practically excludes their spurious interaction. 
For the test case of a single Fe atom on the Ti\textsubscript{2}C surface (Fe/Ti\textsubscript{2}C), preliminary estimations, optimization, and calculation of energy parameters were performed at the $\Gamma$ point for  7x7 and 4x4 Ti\textsubscript{2}C supercells. For the case with the primitive Ti\textsubscript{2}C cell, a denser grid of 5x5 k-points was used.

The adsorption energy $E_a$ was determined according to the formula 
\begin{equation}
 E_a = E_{Fe(Pc)+Ti_2C} - E_{Fe(Pc)} - E_{Ti_2C},
 \label{eq:Adsorption}
\end{equation}
where $Fe(Pc)$ indicates either a single iron atom ($Fe$) or iron phthalocyanine ($FePc$); $Ti\textsubscript{2}C$ - the substrate, and $Fe(Pc)+Ti_2C$ the whole system. To guarantee suitable accuracy of $E_a$, the energies of these three systems are calculated in the same supercell.

To investigate electron transfer, we performed an analysis of the laterally averaged electronic charge density $\rho_{charge} = \rho_{\uparrow} + \rho_{\downarrow}$ and spin density $\rho_{spin} = \rho_{\uparrow} - \rho_{\downarrow}$, where $\rho_{\uparrow}$ and $ \rho_{\downarrow}$ are spin up and spin down electron charge densities, respectively.
The methodology of charge transfer evaluation was inspired by the Bader charge analysis\cite{bader1991quantum}, where the borders between the charge densities of two atoms are determined along the local minimum of the charge densities. Here, we consider the charge transfer between the Ti\textsubscript{2}C surface and the flat FePc molecule or the iron atom. For this purpose, the charge density $\rho$ is integrated over $xy$-planes in each point along the $z$-axis, which gives laterally averaged charge density $\bar{\rho}(z)$
\begin{equation}
\label{eqn:rho_z}
\bar{\rho}(z) = \frac{1}{S} \iint_S \rho(x,y,z) \,dx\,dy,
\end{equation}
where S is the area of the unit cell. A similar approach has been previously implemented to study charge distribution in the hybrid system of VPc on gold surface.\cite{mabrouk2021stability} The minimum of the laterally averaged charge density $\bar{\rho}(z)$ 
between the surface and the molecule could be then considered as a surface (or line) dividing the regions of the substrate and the molecule. The accurate minimum was found using the polynomial approximation in the vicinity of the boundary. 

It is worth noting that the charge density analysis was performed only for valence electrons qualified as such in the employed pseudopotential. The chosen pseudopotentials have the following valence configurations: Ti($3s^24s^23p^63d^2$), C($2s^22p^2$), Fe($4s^13d^7$), N($2s^22p^3$), H($1s^1$). 

\section{Results and disscussion}

\subsection{FePc}

An isolated iron phthalocyanine molecule (Fig. \ref{fig:FePc}) has a tetragonal D\textsubscript{4h} symmetry. The central iron atom is in the square planar ligand field which is created by nitrogen atoms. The strong ligand field makes the iron $d_{x^{2} -y^{2}}$ orbital unfavourable, and, therefore, the ground state of the molecule is triplet. The exact ground state is a topic to study due to the energetic proximity of the two states. While it seems to be commonly accepted that the ground state of FePc is $E_{g}$ with the iron $3d$-shell configuration $d_{xy}^{2} d_{xz}^{2} d_{yz}^{1} d_{z^{2}}^{1} d_{x^{2} -y^{2}}^{0}$, there exist computations predicting $A_{2g}$ state (with the $d_{xy}^{2} d_{xz}^{1} d_{yz}^{1} d_{z^{2}}^{2} d_{x^{2} -y^{2}}^{0}$ iron $3d$-shell configuration) as the ground.\cite{ichibha2017new}

The excited quintet and singlet states can be modelled using one-Slater-determinant DFT methods. Plane-wave methods with the FePc molecule in the cubic cell with the 20 \AA \hspace{0pt} face show that the Fe-N bond lengths are 1.95 \AA\hspace{0pt} in the ground state, 2.01 \AA\hspace{0pt} in the quintet and 1.92 \AA\hspace{0pt} in the singlet. This elongation of the bonds was previously found for molecules with the Fe-N\textsubscript{4} centre.\cite{bhandary2011graphene} The Fe-N bonds lengthening weakens the ligand field created by nitrogen atoms of tetragonal symmetry. Thus, the electrons of the d-shell of the iron atom occupy the higher energy orbitals. The quintet excited state is 0.44 eV higher than the triplet state and the singlet state is 2.68 eV higher.

The FePc symmetry is represented well in the cubic cell. Below, FePc will be studied on the hexagonal lattice structure surfaces. Therefore, the point symmetry of the hybrid system consisting of FePc attached to a substrate is relatively low, the C\textsubscript{2} one. For that reason, optimisation of the geometries of the hybrid systems could lead to FePc geometry slightly distorted from the ideal one. The difference is seen in iron $d$-shell projected densities of states for FePc in cubic and hexagonal lattices (see Supplemental Materials). While in the cubic cell the FePc ground state is $E_g$, in the hexagonal cell it is  $A_{2g}$. These results shed light on the previously mentioned discrepancies of theoretically predicted state of the FePc free standing molecule.

\subsection{Geometric and energy characteristics}

We start the discussion of this hybrid system by presenting the characterisation of its geometry and energetics as obtained in DFT supercell calculations.

\begin{figure}
\includegraphics[width=\columnwidth]{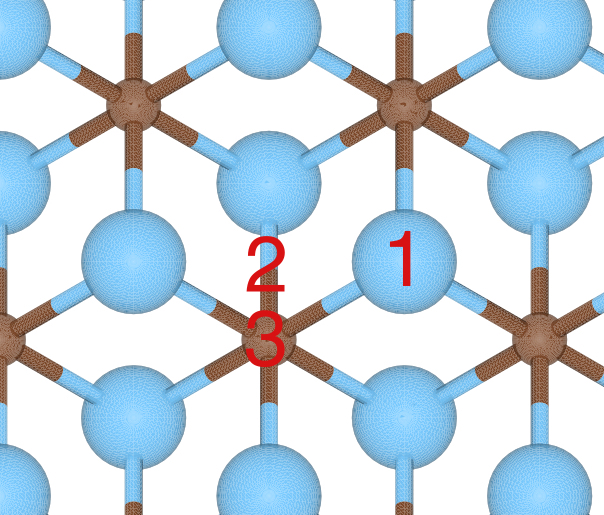}
\caption{Considered points of the FePc adhesion to Ti\textsubscript{2}C. Blue spheres indicate Ti atoms, brown spheres C atoms.
The position of the iron atom is marked with a number: 1) Up-Ti-Centre, 2) Bridge, 3) C-Centre. The figure shows the Ti\textsubscript{2}C layer in the projection on the $xy$-plane. }
\label{fig:Sites}
\end{figure}

The most energetically favourable position of the FePc molecule on the Ti\textsubscript{2}C layer was initially found without taking into account the spin polarisation. 
To accelerate the finding of the optimised geometry of the hybrid system FePc/Ti\textsubscript{2}C, we have performed the following procedures.
We began with constructing a one-dimensional potential energy surface to search for a preliminary distance between the molecule and the surface. The molecule was positioned flat relative to the MXene layer. Based on the calculations performed, further optimisations were carried out at a height of 2.19 \AA.
After that, the positions of the molecule relative to the Ti\textsubscript{2}C layer were selected (Fig. \ref{fig:Sites}), in which the iron atom is located above the upper titanium atom (Up-Ti-Centre), above the carbon atom (C-Centre), and between the upper titanium atoms (Bridge). The position at which the iron atom is located above the lower titanium atom (Down-Ti-Centre) was excluded from consideration, since further optimisation during this case led to the Bridge-type geometry. In the selected positions, a one-dimensional potential energy surface was constructed with the coordinates of the molecule rotation around its axis. After finding the angle with the minimum energy of the system, finally, the geometry of the system was fully optimised by the computer code.

The performed procedure indicates that the Bridge position is the most energetically preferable, with the total energy being 0.13 eV lower than in the Up-Ti-Centre case, and 0.46 eV lower than the total energy of the hybrid system with the C-Center position of FePc. All further calculations were performed only for FePc in the Bridge position.

\begin{figure}
\includegraphics[width=\columnwidth]{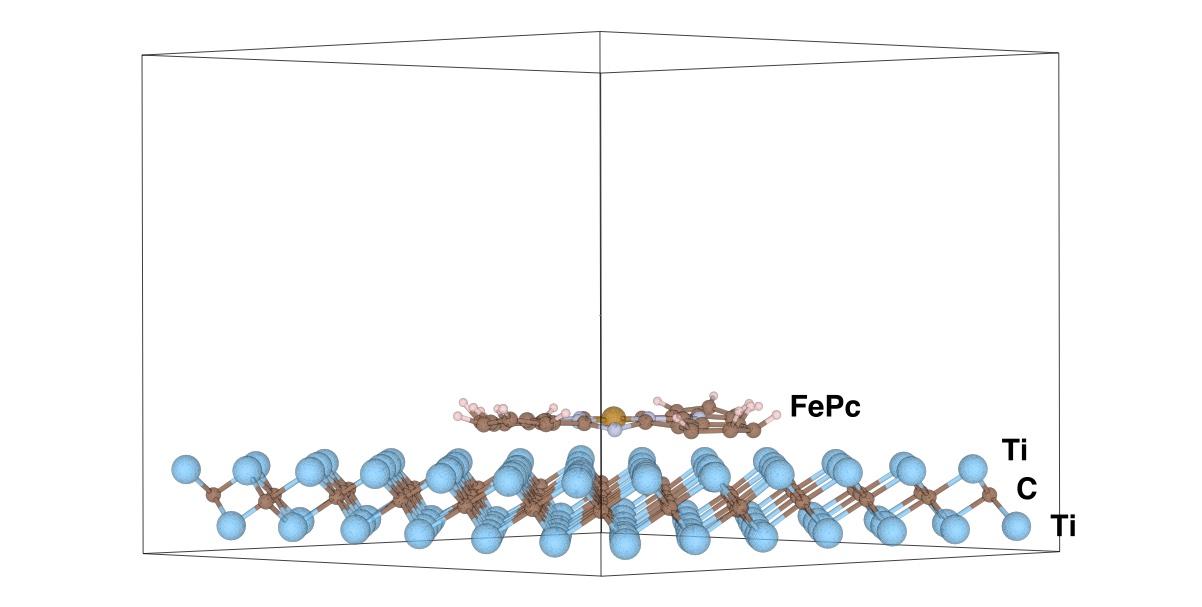}
\caption{Side view of the FePc/Ti\textsubscript{2}C hybrid system in the supercell in the DFT calculations. The vacuum layer in the supercell is in proportion to the lateral dimensions of the supercell.}
\label{fig:FePcTi2C}
\end{figure}

After optimisation (Fig. \ref{fig:FePcTi2C}), the benzene rings of the FePc molecule are slightly bent from the MXnene surface, while the iron atom is located slightly below the average level of the molecule. The average distance between FePc and Ti\textsubscript{2}C is 2.09 \AA, while the Fe-Ti\textsubscript{2}C distance is 2.03 \AA.
The adsorption energy of the molecule on a surface is about 20 eV, which is several times higher than the adsorption of FePc on other two-dimensional surfaces. This value, as well as the small distance between the molecule and the surface, may indicate that the interaction is no longer determined solely by the van der Waals interaction.

We consider also FePc/Ti\textsubscript{2}C hybrid system with FePc molecule in its excited state and observe changes in the geometry of the system in comparison to the case when FePc is in its ground state. 
With the lengthening of the Fe-N bonds, the iron atom noticeably moved out of the plane of the molecule. The average distance between the FePc molecule and the top layer of Ti\textsubscript{2}C is 2.08 \AA, while the distance from the iron atom to Ti\textsubscript{2}C is 2.36 \AA, which is 0.33 \AA\hspace{0pt} longer than in the case of the ground state molecule. This differs from the case of the free molecule, where the bond lengths are lengthened in the plane of the molecule.

\subsection{Polarization cases}

\begin{figure}
\centering
\begin{subfigure}{0.49\columnwidth}
  \centering
  \includegraphics[width=\columnwidth]{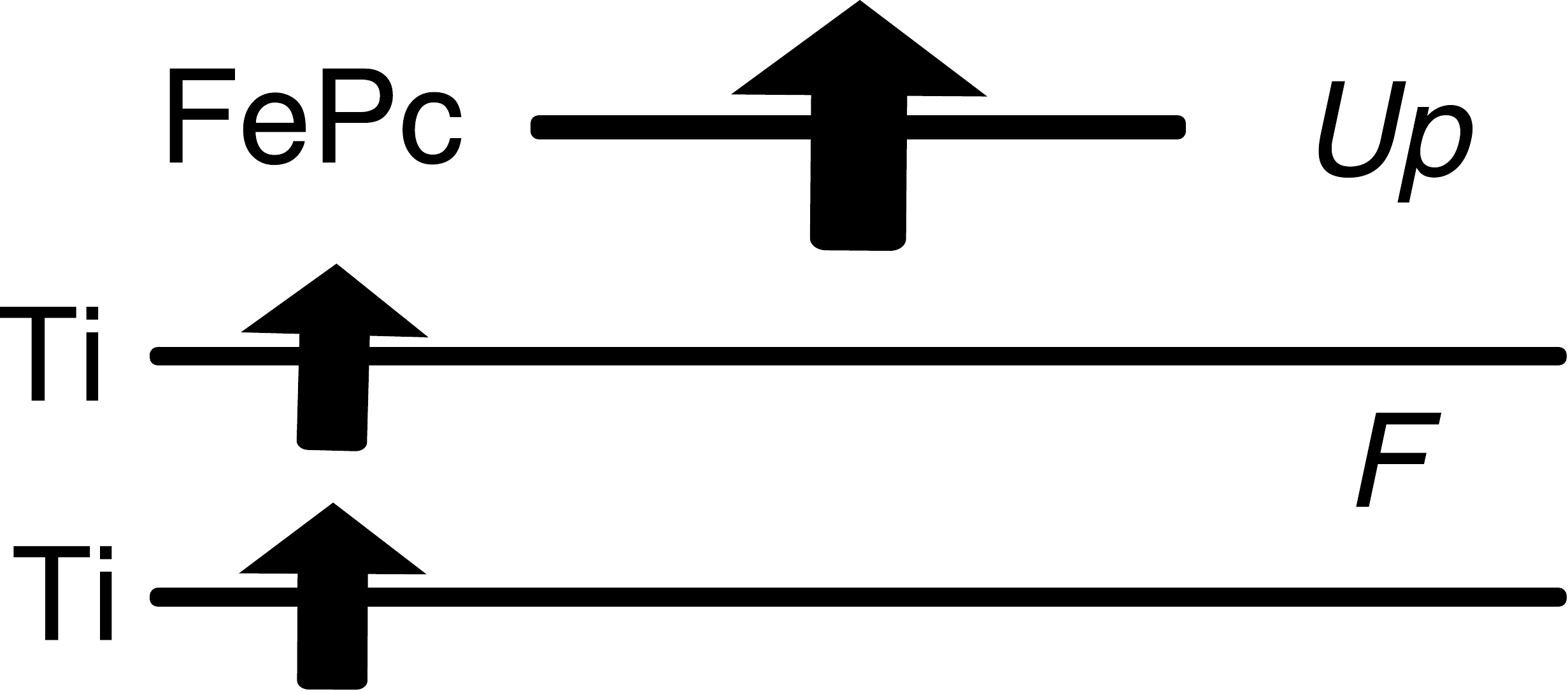}
  \caption{}
  \label{fig:FUp}
\end{subfigure}
\hfill
\begin{subfigure}{0.49\columnwidth}
  \centering
  \includegraphics[width=\columnwidth]{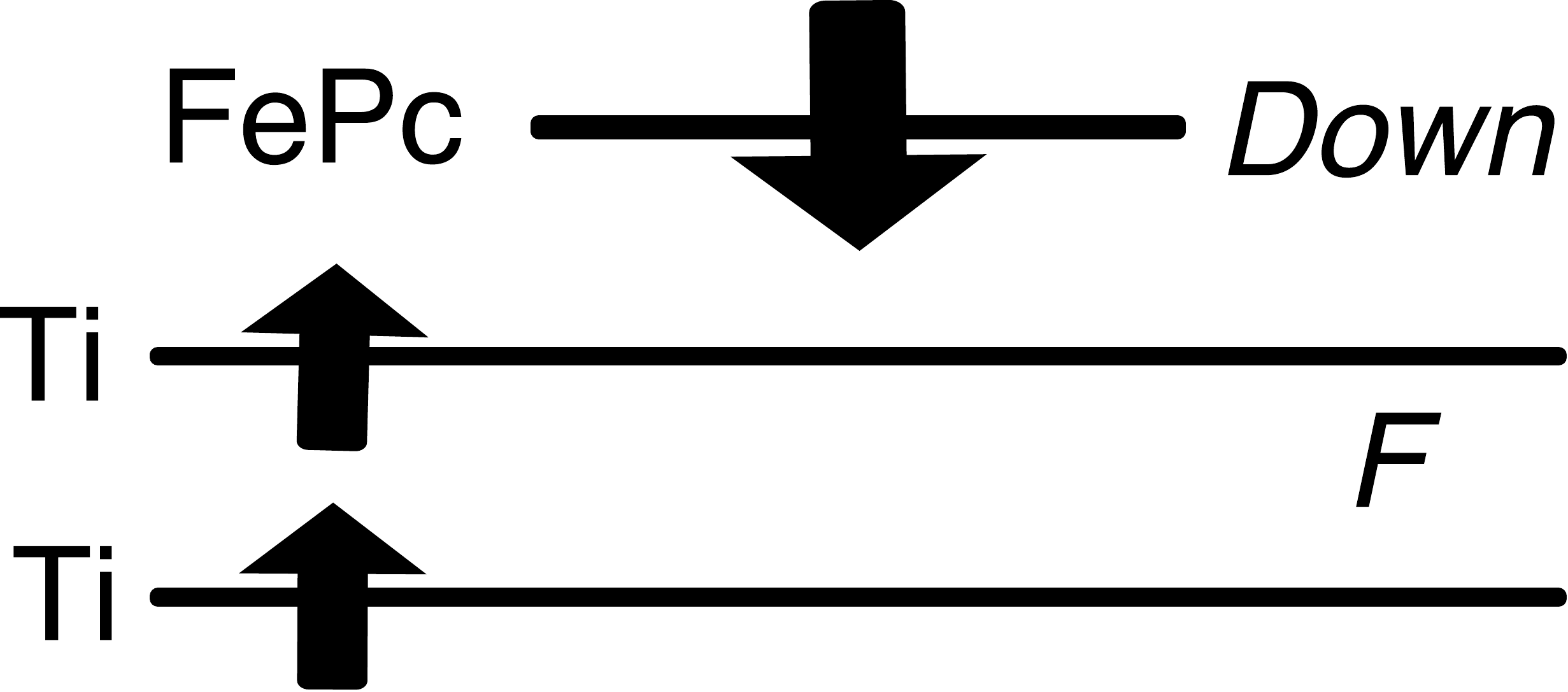}
  \caption{}
  \label{fig:FDown}
\end{subfigure}
\hfill
\begin{subfigure}{0.49\columnwidth}
  \centering
  \includegraphics[width=\columnwidth]{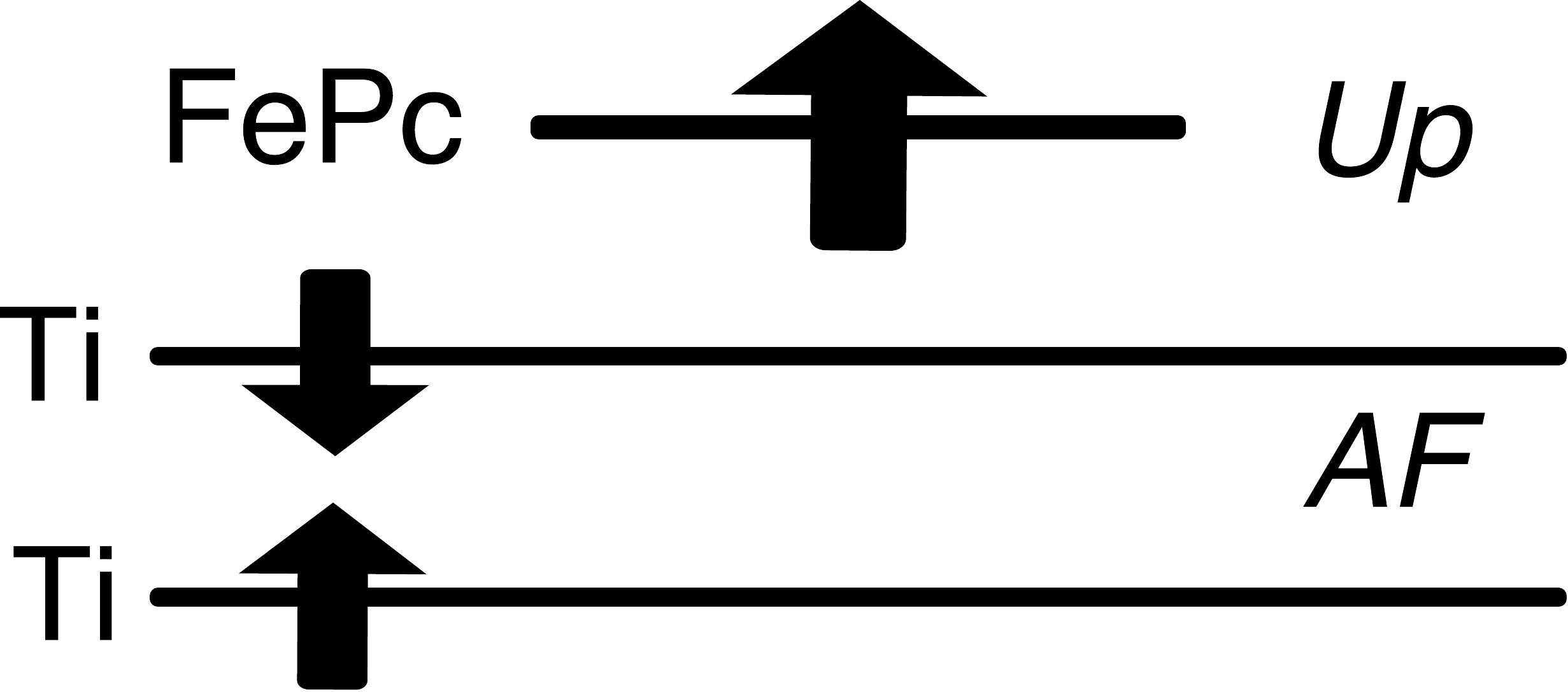}
  \caption{}
  \label{fig:AFUp}
\end{subfigure}
\hfill
\begin{subfigure}{0.49\columnwidth}
  \centering
  \includegraphics[width=\columnwidth]{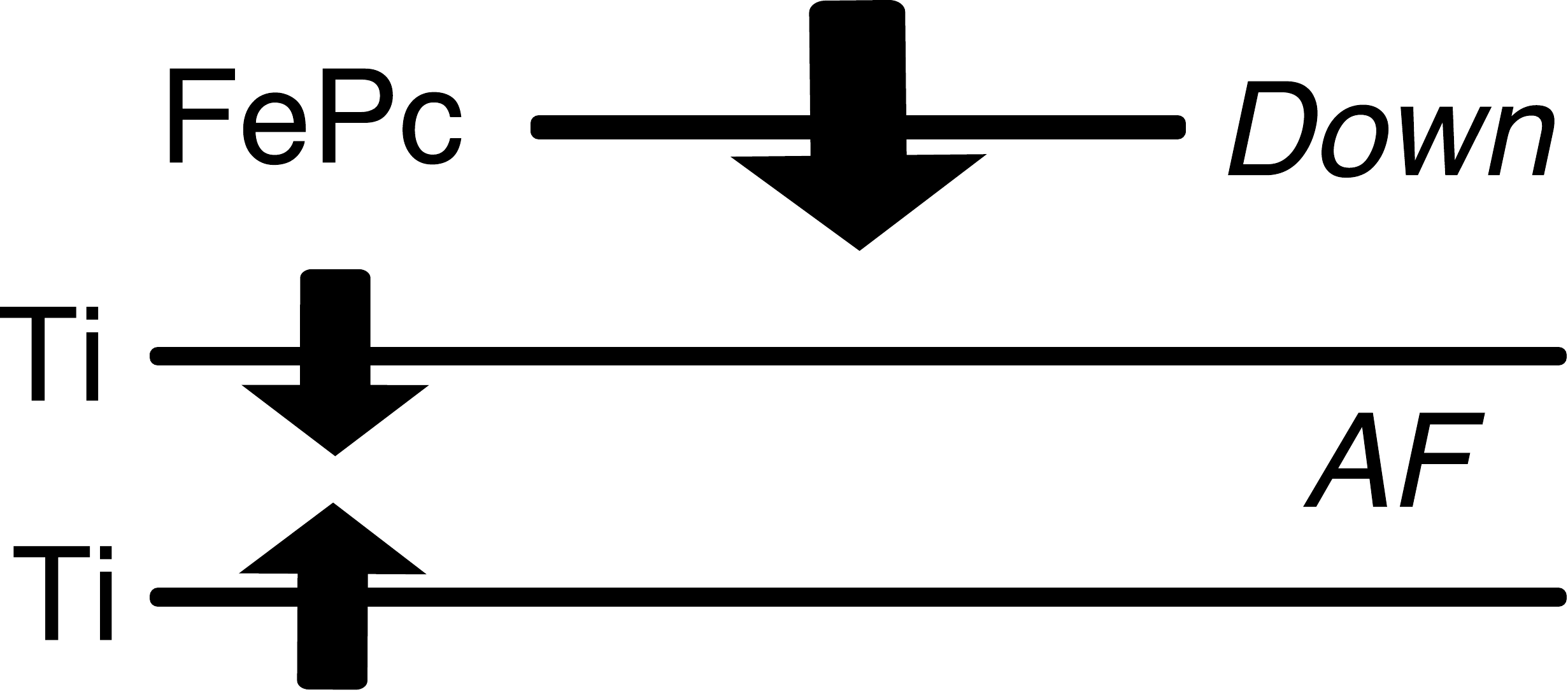}
  \caption{}
  \label{fig:AFDown}
\end{subfigure}
\hfill
\caption{Schematic side view of the possible spin polarisation orientations in FePc molecule and Ti layers indicated as: a) FUp, b) FDown, c)AFUp, d)AFDown}
\label{fig:FePcTi2CPolarisations}
\end{figure}

Taking into account the spin polarisation of this system leads to several possible magnetic configurations. The Ti\textsubscript{2}C surface can be both ferromagnetic and antiferromagnetic, and the FePc molecule can have different directions of its magnetic moment vector relative to Ti\textsubscript{2}C. In total, there are four different magnetic configurations for a given system (Fig. \ref{fig:FePcTi2CPolarisations}). By initialising the initial magnetic moment in this way, the results of calculating the energy parameters showed that the cases with an antiferromagnetic orientation of the titanium layers are energetically more favourable. For the free standing Ti\textsubscript{2}C, the energy difference between the ferromagnetic and antiferromagnetic states is 1.67 eV (for the considered size of the cell). The energies of FePc/Ti\textsubscript{2}C systems with the ferromagnetic orientation of titanium layers turned out to be 0.8 eV higher. The exchange energy $E_ex = E_{Fe_Up} - E_{Fe_Down}$, the energy difference between the cases with the same initial Ti\textsubscript{2}C magnetisation and different orientation of the iron atom in the FePc molecule, is 283 meV for the ferromagnetic case, 9 meV for the antiferromagnetic case, and 0.6 meV for the antiferromagnetic case with the molecule in the excited state.

The magnetic characteristics of the studied systems with the non-excited FePc molecule are shown in Table \ref{table:U4magn}. In the calculations, a feature of the magnetisation of these systems is distinguished. The magnetic moment of a single FePc molecule is 2 $\mu_\text{B}$. When the magnetic moment of a single molecule is reoriented, its value changes by  $|4 \mu_\text{B}|$ from -2$\mu_\text{B}$ to 2$\mu_\text{B}$ or vice versa. However, the table shows that the difference in total magnetisation between the cases of FUp (AFUp) and FDown (AFDown) is only $|2 \mu_\text{B}|$. In this case, the spin density of the FePc molecule is concentrated on the iron atom, and the magnitude of the polarisation of the iron atom does not fundamentally change.

Studying the spin polarisation of atoms (Table \ref{table:U4Polar}) helps to understand the distribution of the magnetic moment in the systems. Atomic spin polarizations were calculated as the difference in the spin density on the atomic orbitals projections. There are cases in which the magnetic moments of the atoms of the upper titanium layer MXene and the iron atom in FePc are initially co-directed (cases FUp and AFDown). In these cases, the titanium atoms closest to the iron atom have practically no magnetic moment. It indicates that a ferromagnetic interaction between the top Ti\textsubscript{2}C layer and the FePc iron atom takes place. The difference in polarisation of titanium atoms adjacent to the iron atom between the cases FUp (AFUp) and FDown (AFDown) nearly compensates for the difference between the sum of atomic polarisation of the entire layer in the same cases.

\begin{table}
    \caption{Magnetic characteristics (in $\mu_\text{B}$/cell) of the FePc/Ti\textsubscript{2}C hybrid system with various polarisation configurations (FUp, FDown, AFUp, and AFDown) as described in the main text and Fig. \ref{fig:FePcTi2CPolarisations} and polarisation configurations AFUp and AFDown with the FePc molecule is in the considered excited state (AFUpEx and AFDownEx).}
    \label{table:U4magn}
\begin{ruledtabular}
\begin{tabular}{ccc}

  & Total Magnetisation & Absolute Magnetisation\\
 \colrule 
 FUp & 67.63 & 72.92 \\
 FDown & 65.42 & 72.65 \\
 AFUp & 23.56 & 81.57 \\
 AFDown & 21.56 & 81.29 \\
  AFUpEx & 25.34  & 82.96 \\
 AFDownEx & 18.31 & 83.89\\

\end{tabular}
\end{ruledtabular}
\end{table}

\begin{table*}
    \caption{The spin polarisations (in a.u.) in the FePc/Ti\textsubscript{2}C hybrid system. The “1st Neighbours Polarisation” column shows the sum polarization of 4 titanium atoms which are the nearest to the FePc iron atom. The “2st Neighbours Polarisation” column shows the sum polarization of 14 top layer titanium atoms which are the nearest to the FePc iron atom.}
    \label{table:U4Polar}
\begin{ruledtabular}
\begin{tabular}{ccccccc}

  & \makecell{Sum of\\Polarisations} & \makecell{Sum of MXene \\Atomic Polarisations}& \makecell{FePc\\Polarisation} & \makecell{Fe\\Polarisation} & \makecell{1st Ti Neighbours\\Polarisaiton} & \makecell{2nd Ti Neighbours\\Polarisaiton} \\
 \colrule 
 FUp & 61.36 & 59.96 & 1.4 & 1.24 & -0.07 & 0.17 \\
 FDown & 59.04 & 60.25 & -1.21 & -1.33 & -0.29 & 0.64 \\
 AFUp & 22.21 & 20.91 & 1.31 & 1.29 & -0.47 & -1.82 \\
 AFDown & 20.36 & 21.52 & -1.16 & -1.33 & -0.04 & -1.16 \\
  AFUpEx & 24.27 & 21.17 & 3.1 & 2.93 & -0.34 & -1.67 \\
 AFDownEx & 16.55 & 20.37 & 3.82 & -3.61 & -0.47 & -1.83 \\
\end{tabular}
\end{ruledtabular}
        \end{table*}

The magnetic characteristics of FePc / Ti\textsubscript{2}C with the excited state molecule are also presented in Table \ref{table:U4magn}. Only the cases with the antiferromagnetic polarization Ti\textsubscript{2}C were studied (labeled as AFUpEx and AFDownEx), since they have lower energy compared to the cases with the ferromagnetic polarization. The unobvious difference in the complete magnetization of systems with opposite orientations of the spin moment of the Fe atom, which was observed in the case with the ground state molecule, persisted in the case of the excited state. The difference between the excited states (S = 2, $\mu$ = 4 $\mu_\text{B}$) of the free standing FePc molecule with antiparallel spin orientations is $|8 \mu_\text{B}|$, while for FePc/Ti\textsubscript{2}C with antiparallel FePc spin orientations it is $|7 \mu_\text{B}|$.

The results of the spin density projection onto the atomic orbitals for the FePc/Ti\textsubscript{2}C hybrid systems when the FePc molecule is in the excited state
are presented in Table \ref{table:U4Polar}. 
Here, the "1st Ti Neighbours" atoms adjacent to the iron atom have spin polarisation if the iron atom and the Ti upper layer magnetisations are co-directed. For the same system with FePc in the ground state it is shown that the "1st Ti Neighbours" spin polarization is practically absent.
The described difference between systems in ground and excited FePc states can be explained by the fact that in the system with excited FePc the iron atom is above the FePc plane and, as a consequence, the interaction between the iron atom and the upper Ti\textsubscript{2}C layer decreases.

\subsection{Charge Transfer Analysis}

\begin{figure*}
\centering
\begin{subfigure}{\columnwidth}
  \centering
  \includegraphics[width=\linewidth]{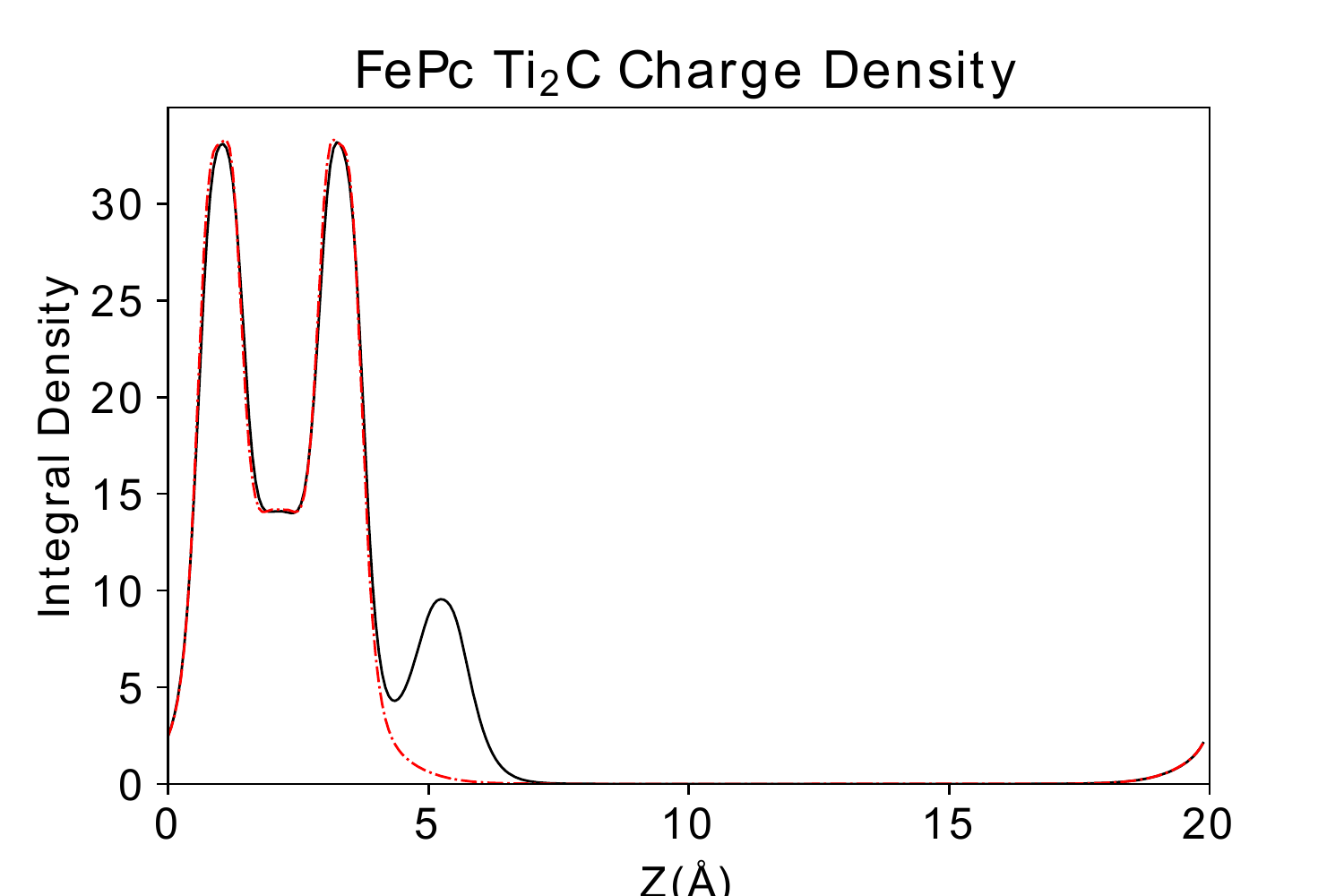}
  \subcaption{}
  \label{fig:FePcTi2CChDensFull}
\end{subfigure}
\hfill
\begin{subfigure}{\columnwidth}
  \centering
  \includegraphics[width=\linewidth]{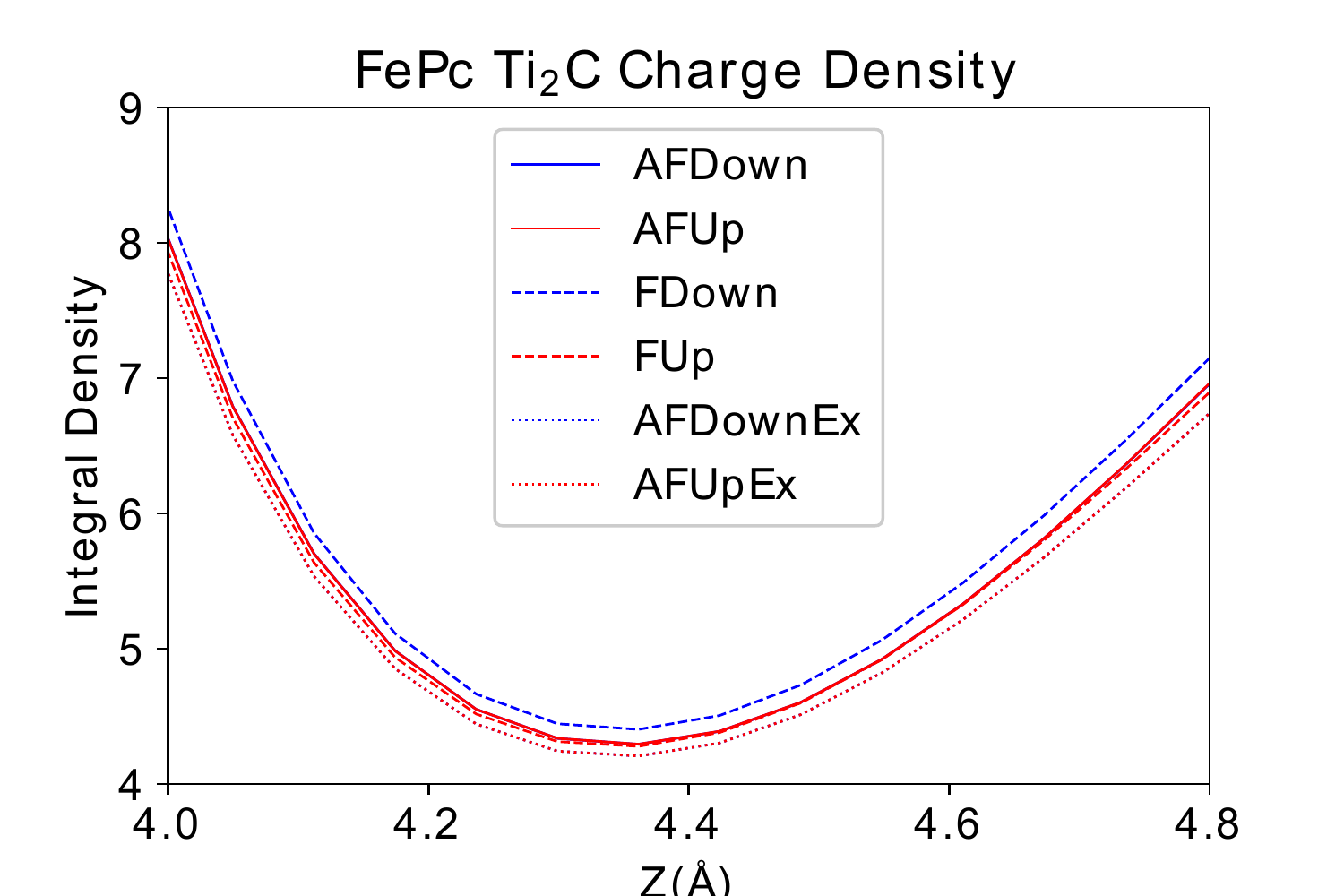}
  \subcaption{}
  \label{fig:FePcTi2CChDensPrecise}
\end{subfigure}
\caption{The laterally averaged charge density of valence electrons (in units of 1/\AA\textsuperscript{3}) for the six types of the FePc/Ti\textsubscript{2}C hybrid system: (a) black -  the FePc/Ti\textsubscript{2}C charge density, red dashed -  the Ti\textsubscript{2}C charge density; (b) close-up graph in the vicinity of minimum for all considered systems.}
\label{fig:FePcTi2CChDens}
\end{figure*}

The plots of the $xy$-integrated charge density $\bar{\rho}_{charge}$ for FePc/Ti\textsubscript{2}C are shown in Fig. \ref{fig:FePcTi2CChDens}. The total charge density integral over the unit cell was equal to the number of valence electrons in the cell. It is worth noting that the view of plots is practically identical for all considered types of hybrid systems, and the difference is not visible on the original scale. The charge densities of all systems near the Bader minimum are shown in the Fig. \ref{fig:FePcTi2CChDensPrecise}. The 6-th order polynomial fitting was used to find the minimum for all configurations, and the charge transfer to the molecule is shown in Table \ref{table:BaderChTransfer}. 

\begin{table}
\caption{Charge transfer (in electrons) to the FePc molecule in the FePc/Ti\textsubscript{2}C system with various magnetisation configurations.}
        \label{table:BaderChTransfer}
        \centering
\begin{ruledtabular}
\begin{tabular}{cc}
 System & Charge transfer to FePc \\
\colrule
 AFDown & 8.45 \\
 AFUp & 8.45 \\
 FDown & 8.4 \\
 FUp & 8.77 \\
 AFDownEx & 8.62 \\
 AFUpEx & 8.74 \\
\end{tabular}
\end{ruledtabular}
        \end{table}


To investigate the target of the charge transfer from Ti\textsubscript{2}C to FePc the analysis of the projected density of states was done. This approach does not show the precise results of the charge transfer but can describe the Löwdin charge \cite{lowdin1950non} difference for each atom in the FePc molecule. In Table \ref{table:LöwdinChTransfer}, there are given mean and overall valence orbital charges for each type of atom in FePc. The results are shown for the free FePc molecule and for FePc in the FePc/Ti\textsubscript{2}C hybrid system. 
The last column shows the mean charge transfer per atom (the difference between the atomic Löwdin charge in FePc/Ti\textsubscript{2}C and FePc). This demonstrates that charge transfer occurs predominantly to carbon atoms. 

\begin{table*}
\caption{Löwdin charge for valence electrons in the free and adsorbed FePc molecule for each type of atom and the average charge transfer per atom for each type.}
        \label{table:LöwdinChTransfer}
         \centering
\begin{ruledtabular}
\begin{tabular}{cccc}
  & Löwdin charge in pure FePc & Löwdin charge in FePc/Ti\textsubscript{2}C & Charge transfer per atom \\
\colrule
 C & Mean = 3.971 (Sum = 127.084) & 4.147 (132.713) & 0.176 \\
 Fe & 7.075 & 7.454 & 0.379 \\
 N & 5.283 (42.266) & 5.356 (42.849) & 0.073 \\
 H & 0.817 (13.068) & 0.81 (12.962) & -0.006 \\
\end{tabular}
\end{ruledtabular}
        \end{table*}

\begin{figure*}
\centering
\begin{subfigure}{\columnwidth}
  \centering
  \includegraphics[width=\linewidth]{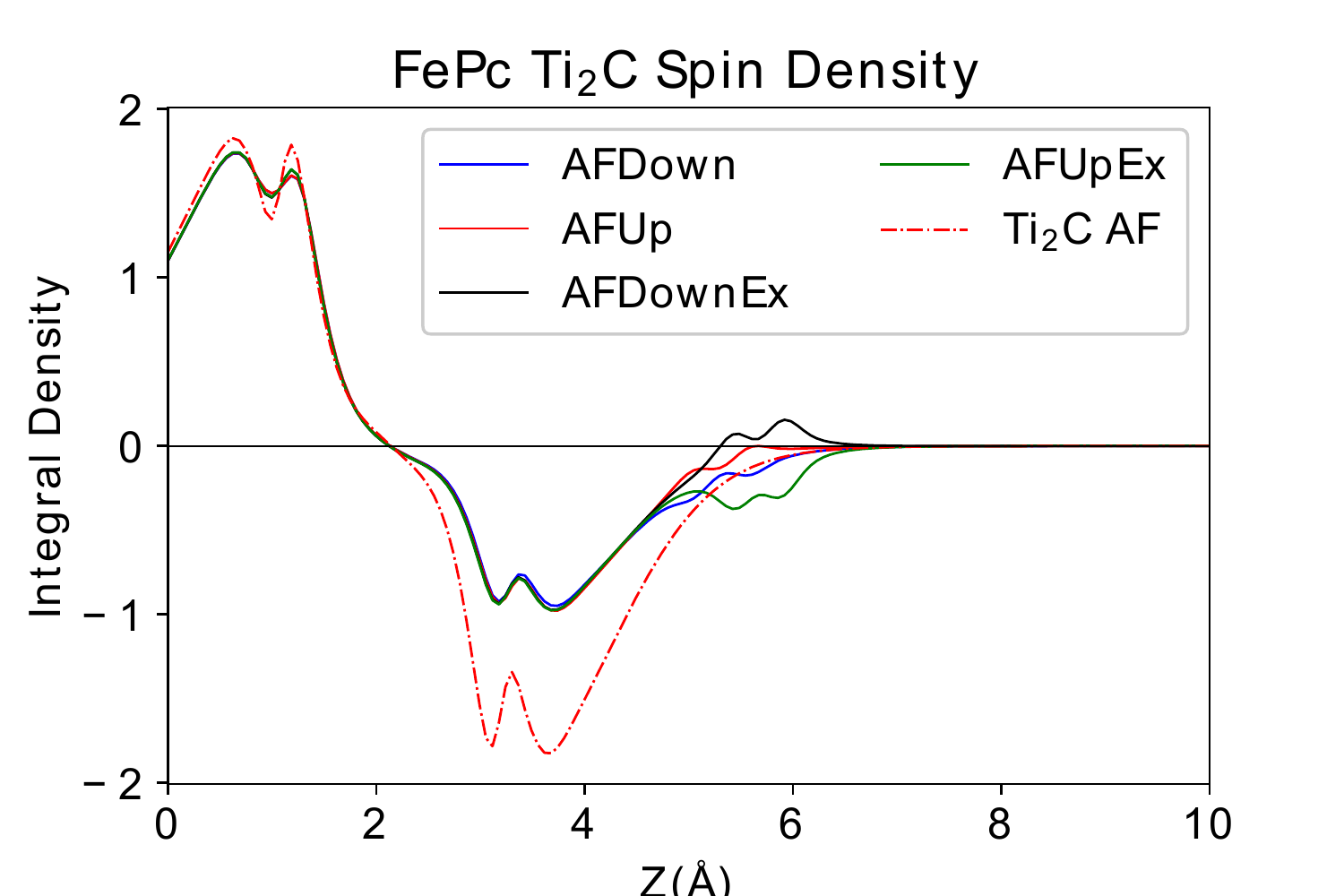}
  \subcaption{}
  \label{fig:FePcTi2CAFSpDens}
\end{subfigure}
\hfill
\begin{subfigure}{\columnwidth}
  \centering
  \includegraphics[width=\linewidth]{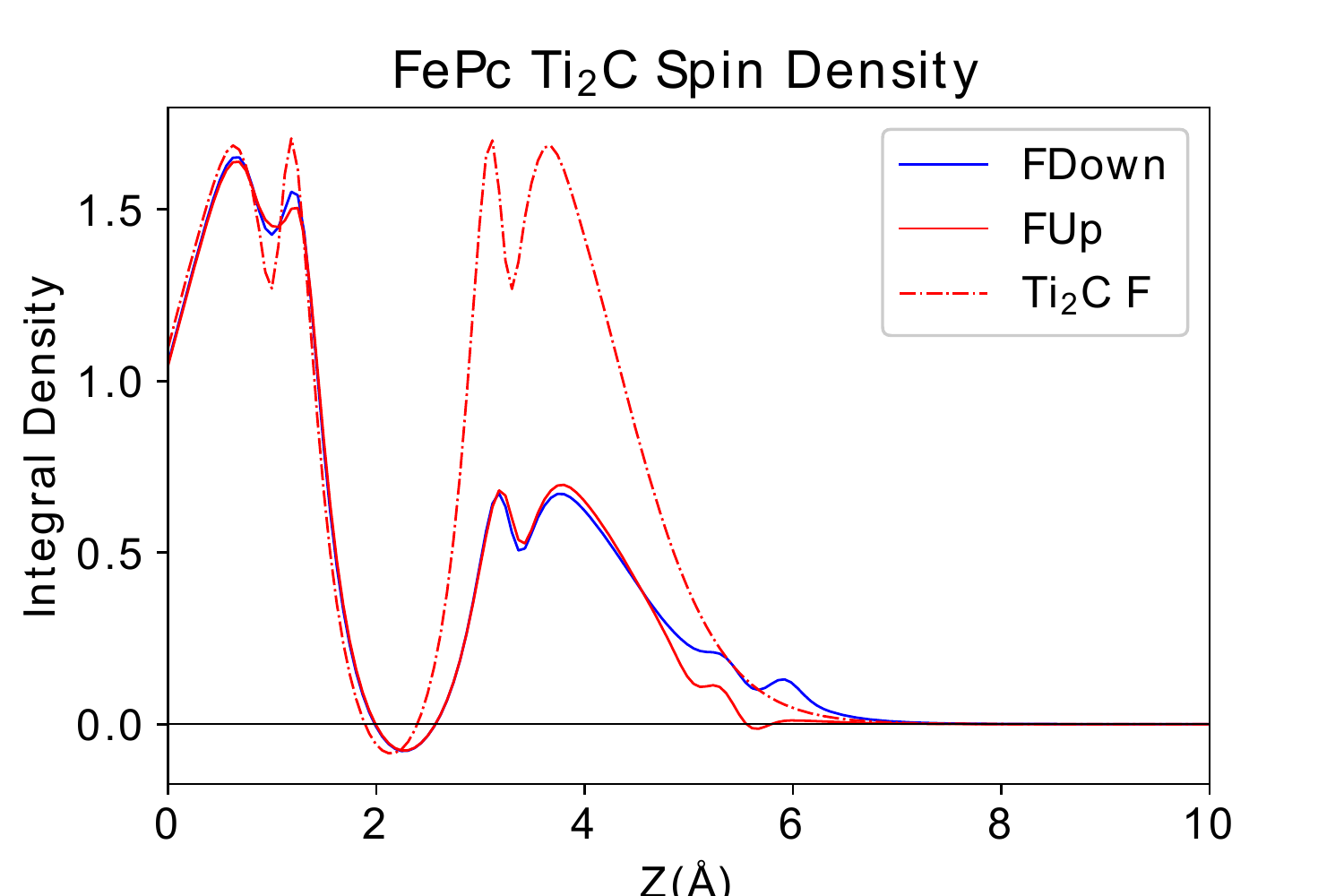}
  \subcaption{}
  \label{fig:FePcTi2CFSpDens}
\end{subfigure}
\caption{The laterally averaged spin density of valence electrons (in units of 1/\AA\textsuperscript{3}) for the FePc/Ti\textsubscript{2}C hybrid systems in the (a) antiferromagnetic and (b) ferromagnetic Ti\textsubscript{2}C configurations. The laterally averaged spin density for Ti\textsubscript{2}C  trilayer with AF and F ordering of magnetic moments in Ti layers is indicated, respectively, in panel (a) and (b) through red-dashed line.}
\label{fig:FePcTi2CSpDens}
\end{figure*}

The laterally averaged, i.e., $xy$-integrated, spin densities $\bar{\rho}_{spin}$ of the FePc/Ti\textsubscript{2}C hybrid systems are shown in Figs. \ref{fig:FePcTi2CAFSpDens} and \ref{fig:FePcTi2CFSpDens}, and compared to the laterally averaged spin densities for Ti\textsubscript{2}C AF and F ordering of magnetisation in the Ti layers. The spin density distributions in FePc/Ti\textsubscript{2}C show a big decline in the upper-Ti layer. Also, two maximums in  the vicinity of both titanium layers show the dominance of non-$xy$-plane $d$-orbitals (${d_{xz}}, d_{yz}, d_{z^2}$) in the MXene formation. 

The spin polarization around FePc also splits into two maxima. In comparison, the spin density distribution for the free FePc molecule has a strong maximum in the centre of the molecule. This indicates that redistribution of spin polarisation occurred in the FePc molecule when it was attached to the MXene surface. Table \ref{table:FeSpPol} shows quantitative changes of iron $d$-orbital spin polarisation. This shows that the polarisation is shifting from $d_{x^{2} -y^{2}}$ and $d_{xy}$ orbitals to the non-plane orbitals.

\begin{table}
\caption{Spin polarization of iron d-orbitals in the free-standing FePc and the  FePc/Ti\textsubscript{2}C hybrid system (AFDown case) }
\label{table:FeSpPol}
\centering
\begin{ruledtabular}
\begin{tabular}{ccc}
  & Fe in FePc & Fe in FePc/Ti\textsubscript{2}C (AFDown) \\
\colrule 
 $d_{z^2}$ & 0.94 & -0.25 \\
 $d_{xz}$ & 0.06 & -0.08 \\
 $d_{yz}$ & 0.05 & -0.89 \\
 $d_{x^{2} -y^{2}}$ & 0.70 & -0.09 \\
 $d_{xy}$ & 0.40 & -0.01 \\
\end{tabular}
\end{ruledtabular}
        \end{table}

\subsection{Fe/Ti\textsubscript{2}C and H\textsubscript{2}Pc/Ti\textsubscript{2}C Analysis}

\begin{figure}
  \centering
  \includegraphics[width=\columnwidth]{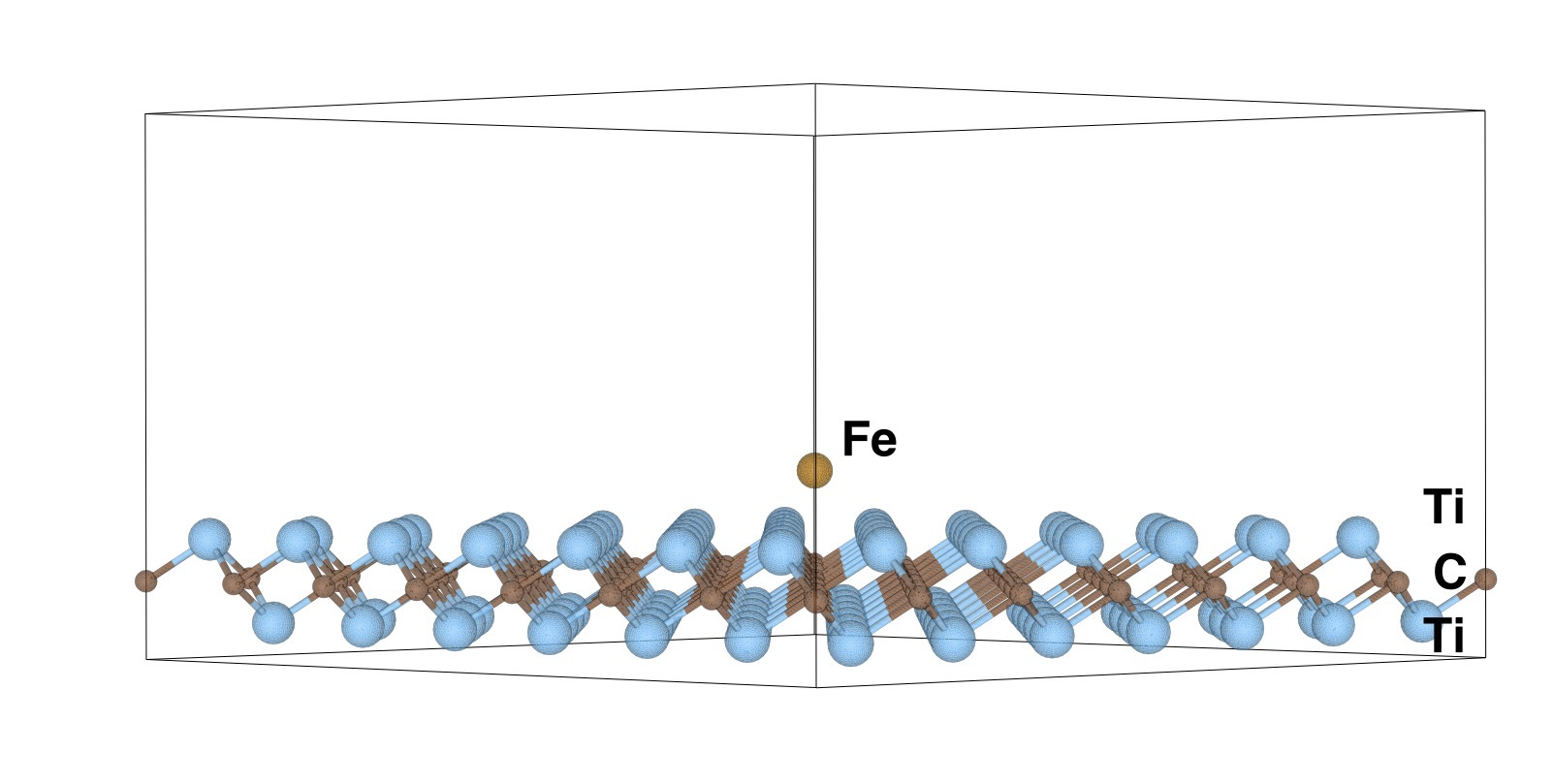}
  \caption{Side view of the Fe/Ti\textsubscript{2}C system}
  \label{fig:FeTi2CView}
\end{figure}

\begin{figure}
\centering

\begin{subfigure}{0.49\linewidth}
  \centering
  \includegraphics[width=\linewidth]{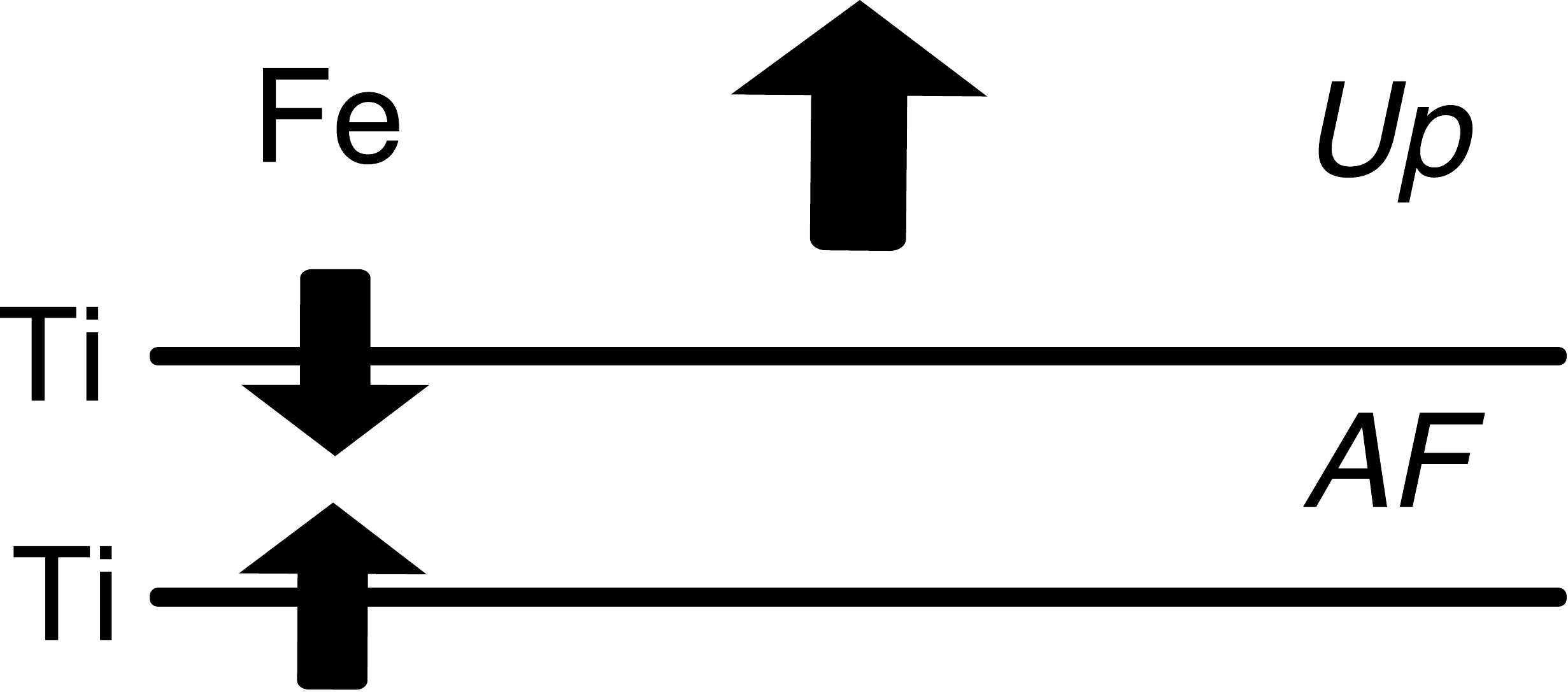}
  \subcaption{}
  \label{fig:FeTi2CUpScheme}
\end{subfigure}
\hfill
\begin{subfigure}{0.49\linewidth}
  \centering
  \includegraphics[width=\linewidth]{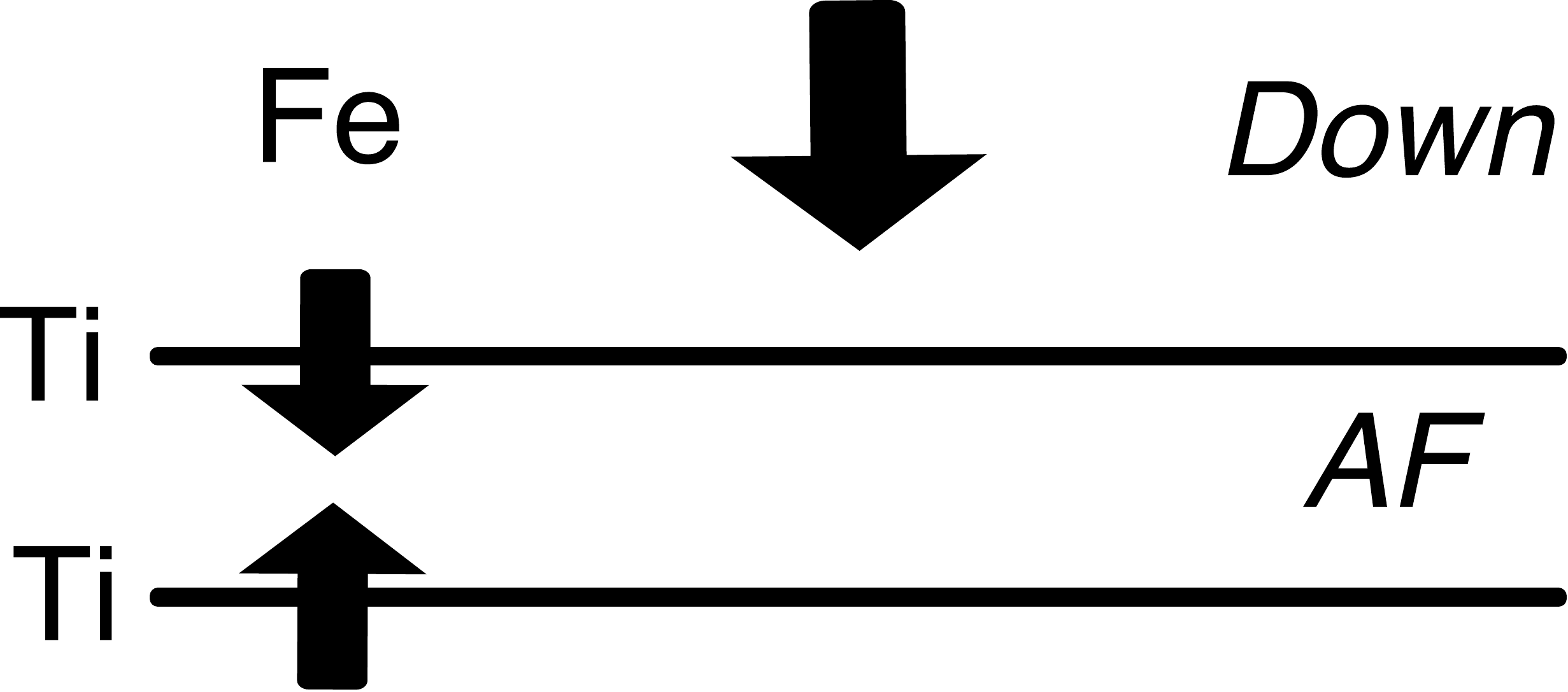}
  \subcaption{}
  \label{fig:FeTi2CDownScheme}
\end{subfigure}

\caption{Schematic side view of the magnetic moments ordering in the  Fe/Ti\textsubscript{2}C structures: (a) FeUp, (b) FeDown cases}
\label{fig:FeTi2CSchemes}
\end{figure}

Previously we described the magnetic interaction in the vicinity of the iron atom in the FePc/Ti\textsubscript{2}C hybrid system. It was found that the iron's spin polarisation influences spin polarisations of neighbouring titanium atoms. A simpler model system with iron atom on the top of the Ti\textsubscript{2}C surface (Fig. \ref{fig:FeTi2CView}) can show this magnetic interaction clearer (without distortions induced by the FePc phthalocyanine ligand).
The calculations were done only for the antiferromagnetic Ti\textsubscript{2}C configuration where the iron spin moment is directed towards Ti\textsubscript{2}C (FeUp, Fig. \ref{fig:FeTi2CUpScheme}) and in the opposite direction (FeDown, Fig. \ref{fig:FeTi2CDownScheme}).


\begin{figure*}
\centering
\begin{subfigure}{0.49\linewidth}
  \centering
  \includegraphics[width=\linewidth]{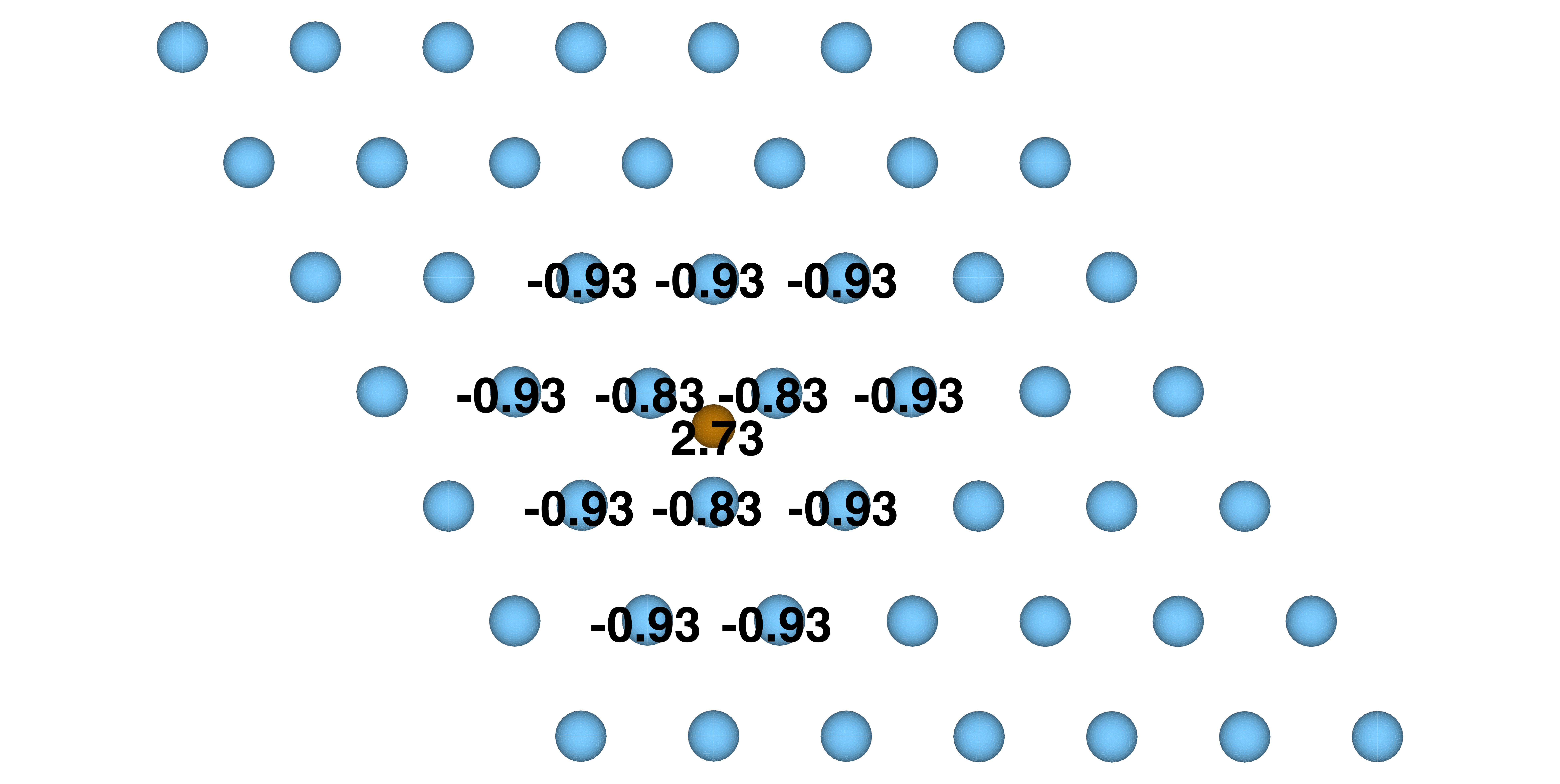}
  \caption{}
  \label{fig:FeTi2C7x7UpPolarisation}
\end{subfigure}
\hfill
\begin{subfigure}{0.49\linewidth}
  \centering
  \includegraphics[width=\linewidth]{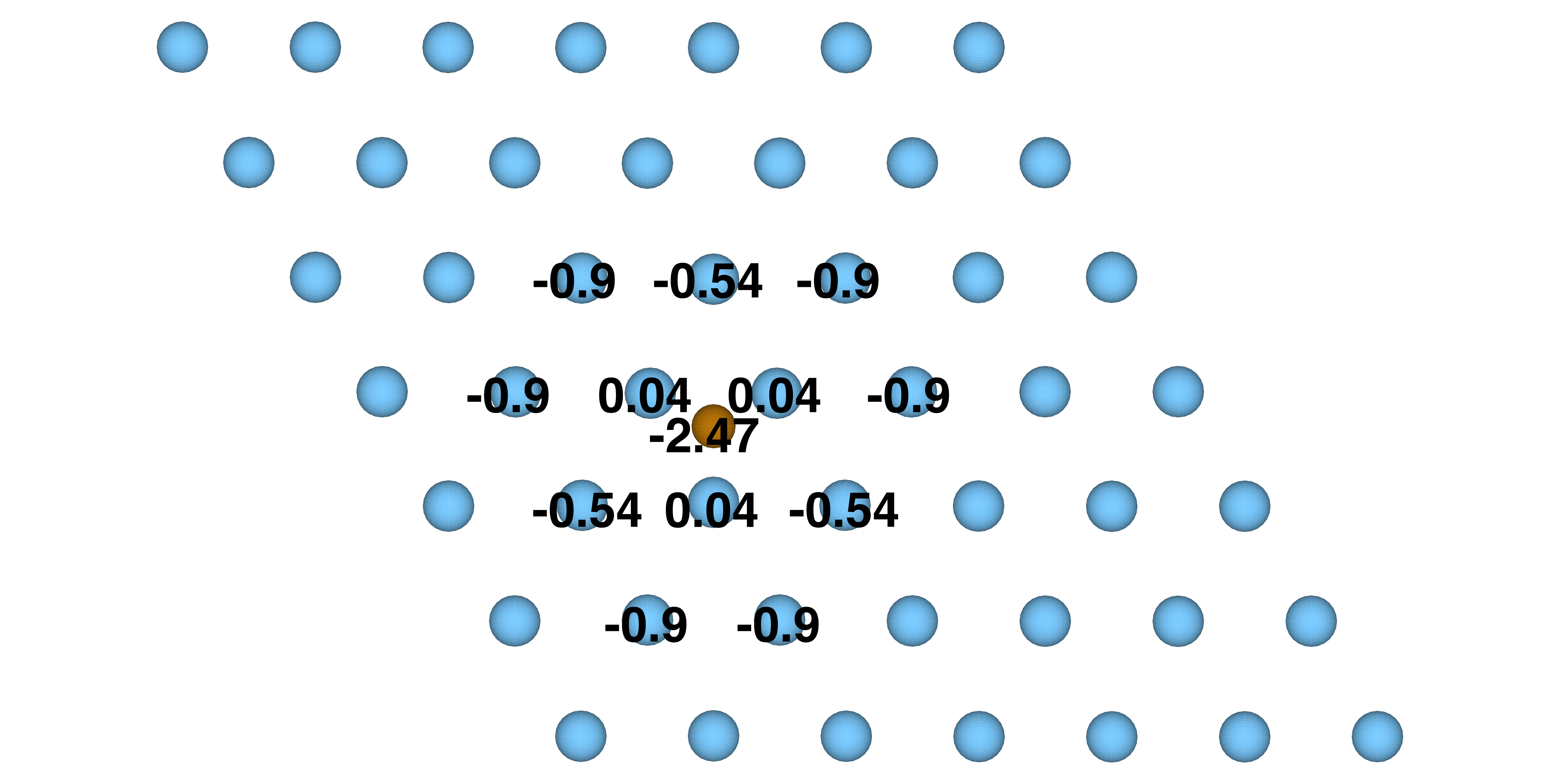}
  \caption{}
  \label{fig:FeTi2C7x7DownPolarisation}
\end{subfigure}
\begin{subfigure}{0.49\linewidth}
  \centering
  \includegraphics[width=\linewidth]{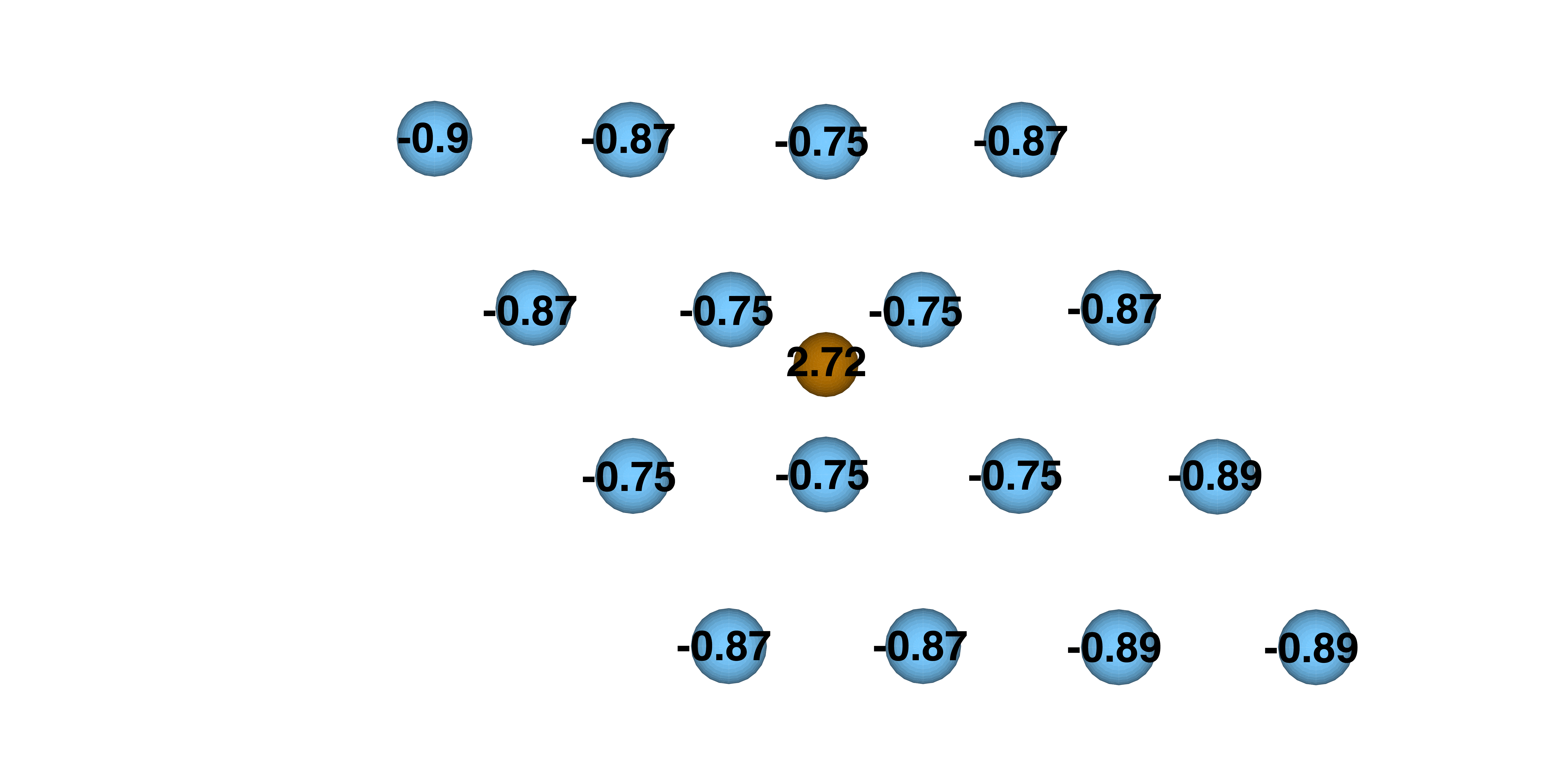}
  \caption{}
  \label{fig:FeTi2C4x4UpPolarisation}
\end{subfigure}
\begin{subfigure}{0.49\linewidth}
  \centering
  \includegraphics[width=\linewidth]{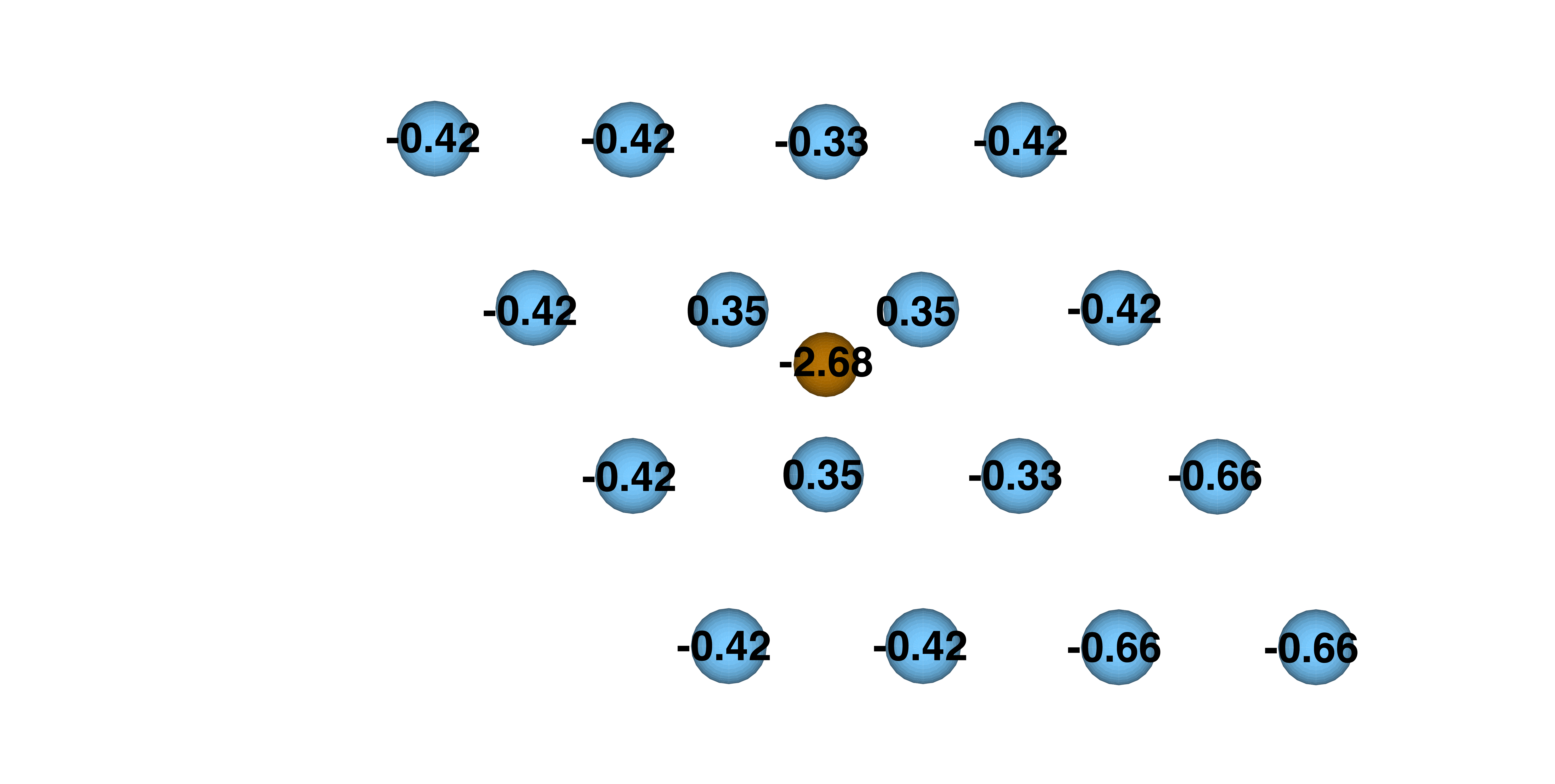}
  \caption{}
  \label{fig:FeTi2C4x4DownPolarisation}
\end{subfigure}
\caption{Spin polarisations of the iron atom and first layer Ti atoms in the Fe/Ti\textsubscript{2}C system with the 7x7 Ti\textsubscript{2}C supercell ((a) FeUp case, (b) FeDown case) and the 4x4 Ti\textsubscript{2}C supercell ((c) FeUp case, (d) FeDown case)}
\label{fig:FeTi2CPolarisations}
\end{figure*}

The computations for the Fe/Ti\textsubscript{2}C system were done with the same 7x7 Ti\textsubscript{2}C supercell, which was used in the studies of FePc/Ti\textsubscript{2}C systems. After preliminary optimization, it was found that the configuration in which the iron atom is located above the middle of the triangle with titanium atoms of the upper layer at the vertices has the lowest energy. With this initial position, the optimization of the iron atom on the antiferromagnetic Ti\textsubscript{2}C layer was performed. The results of calculations for two cases in which the spin moment of the iron atom is directed towards and away from Ti\textsubscript{2}C are presented in Table \ref{table:FeTi2CCharact}. Based on the data obtained, the upward orientation of the iron atom spin is more stable, although the distance of the atom to the layer increases. With this orientation, the iron atom practically does not influence the polarization of the Ti\textsubscript{2}C atoms. In the case of the downward orientation, the polarization of titanium atoms -  the nearest neighbours of the iron atom - is practically zero. This case is similar to the situation with the considered FePc molecule on Ti\textsubscript{2}C when the magnetic orientations of FePc and the upper Ti\textsubscript{2}C layer coincide. But in the case of a single iron atom, its second-row neighbours already have spin polarization comparable to the rest of the titanium atoms in Ti\textsubscript{2}C. The graphic results of the Löwdin analysis are presented in Figs. \ref{fig:FeTi2C7x7UpPolarisation} and \ref{fig:FeTi2C7x7DownPolarisation}.

\begin{table*}
\caption{Geometric, electronic and magnetic characteristics of Fe/Ti\textsubscript{2}C system in different cells}
\label{table:FeTi2CCharact}
        \centering
\begin{ruledtabular}
\begin{tabular}{ccccccc}
  &  & Adsorption Energy, eV & Fe height from Ti\textsubscript{2}C, \AA & $\mu$(Fe) & mean $\mu$(Ti\textsubscript{bottom}) & $\mu$(Ti\textsubscript{Fe neigbours}) \\
\colrule
 \multirow{2}{*}{7x7 Ti\textsubscript{2}C supercell} & FeUp & -4.02 & 1.83 & 2.73 & 0.95 & -0.82 \\
   & FeDown & -3.59 & 1.77 & -2.47 & 0.94 & 0.03 \\
 \multirow{2}{*}{4x4 Ti\textsubscript{2}C supercell} & FeUp & -5.87 & 1.79 & 2.72 & 0.93 & -0.75 \\
   & FeDown & -5.74 & 1.76 & -2.67 & 0.76 & 0.35 \\
 \multirow{2}{*}{Primitive Ti\textsubscript{2}C cell} & FeUp & -2.14 & 2.07 & 3.01 & 1.32 & -0.59 \\
   & FeDown & -2.09 & 2.08 & -2.96 & 0.76 & 0.42 \\
\end{tabular}
\end{ruledtabular}
        \end{table*}


The calculations with higher concentrations of iron atoms on the Ti\textsubscript{2}C surface were done to compare the energetic and magnetic characteristics. In the cases where 4x4 Ti\textsubscript{2}C supercell is used, the adsorption energy is about one and a half times higher but the atom-layer distance is similar to the 7x7 supercell case. A remarkable fact is that the magnetic moment on the titanium atoms that one the neighbours of the iron atom, does not decay as in the 7x7 supercell case but changes its sign to the opposite. Also, the decrease of the bottom Ti layer spin density is observed in the FeDown case. Spin polarisations of the iron atom and top Ti\textsubscript{2}C layer atoms are shown in Fig. \ref{fig:FeTi2C4x4UpPolarisation} for the FeUp case and Fig. \ref{fig:FeTi2C4x4DownPolarisation} for the FeDown case.


The Fe/Ti\textsubscript{2}C complex with very high concentration of iron atoms, where an iron atom is on the top of each atom was also considered. In principle, here iron atoms are so close to each other that they form the additional sublayer above MXene. The adsorption energy of the atom to the layer is two times lower and the atom-layer distance is 0.3 \AA\hspace{0pt} longer. In the FeDown case, the upper titanium layer magnetisation was flipped from down to up what changed the Ti\textsubscript{2}C configuration  from antiferromagnetic to ferromagnetic.

\begin{figure*}
\centering
\begin{subfigure}{\columnwidth}
  \centering
  \includegraphics[width=\linewidth]{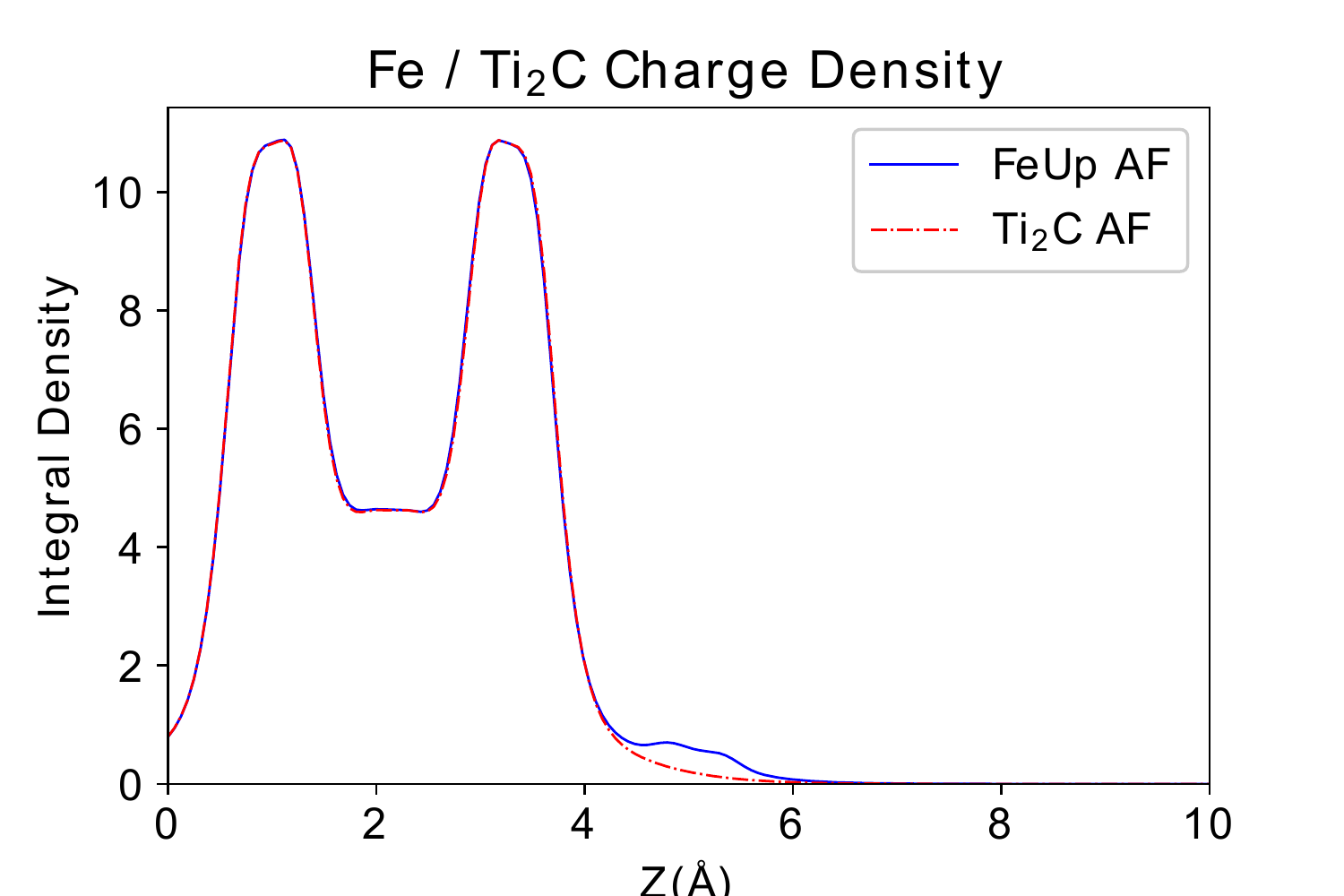}
  \subcaption{}
  \label{fig:FeTi2CChDens}
\end{subfigure}
\hfill
\begin{subfigure}{\columnwidth}
  \centering
  \includegraphics[width=\linewidth]{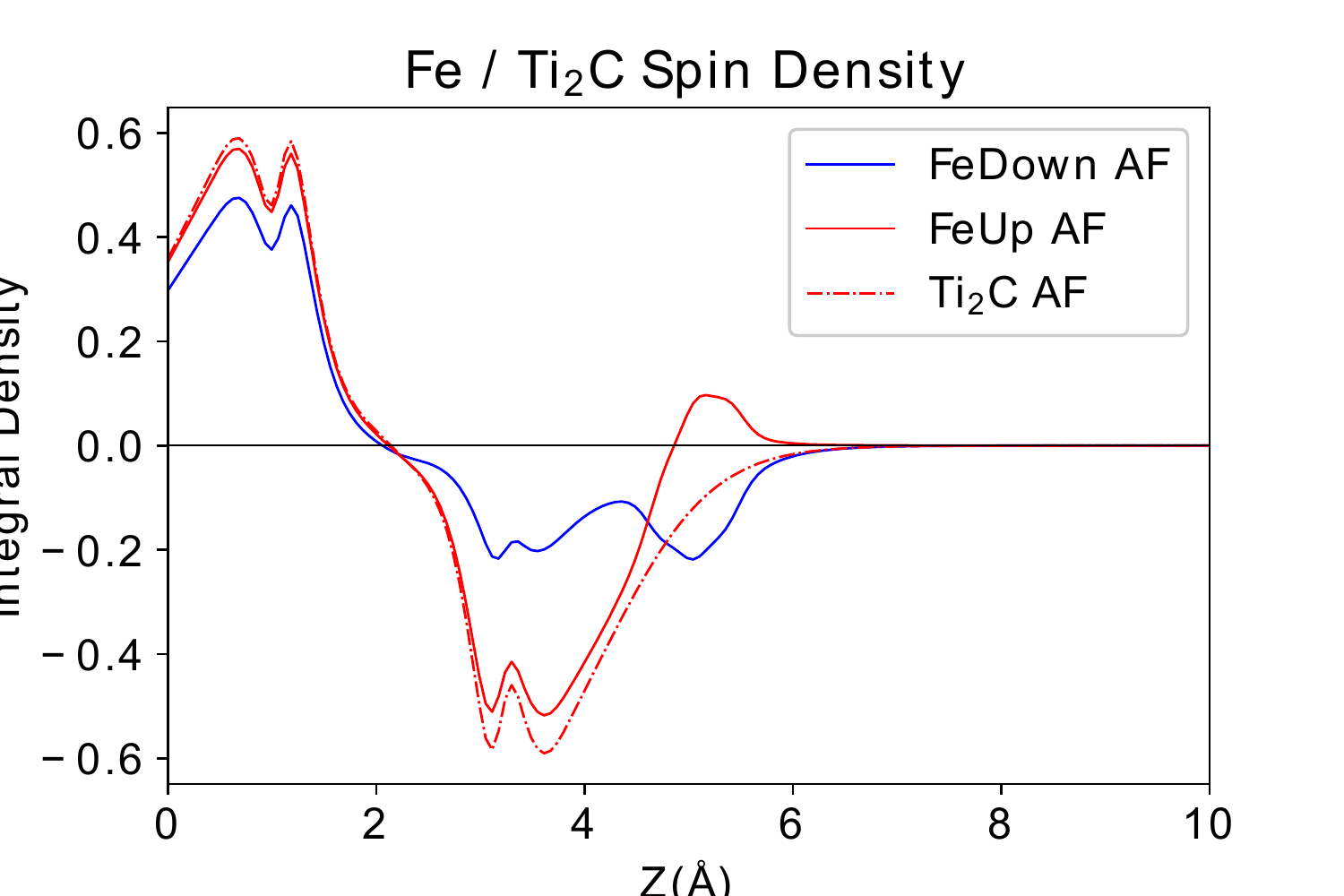}
  \subcaption{}
  \label{fig:FeTi2CSpDens}
\end{subfigure}
\caption{The laterally averaged (a) charge and (b) spin densities of valence electrons (in units of 1/Å\textsuperscript{3}) for the Fe/Ti\textsubscript{2}C system.}
\label{fig:FeTi2CChSpDens}
\end{figure*}

The electron density analysis of the Fe/Ti\textsubscript{2}C system was done using the same techniques as in the case of FePc/Ti\textsubscript{2}C hybrid system. 
The analysis was done for the system with one iron atom in the 4x4 MXene supercell. 
The vertical distributions of charge and spin densities are depicted in Fig. \ref{fig:FeTi2CChSpDens}. Despite the fact that the iron atom does not really change the charge density distribution (Fig. \ref{fig:FeTi2CSpDens}), it does influence the spin density distribution (Fig. \ref{fig:FeTi2CChDens}). The spin density distribution illustrates well the influence of the iron atom on its titanium neighbours; the issue that has been discussed in previous reports. The  iron "up" spin polarisation has almost no effect on the MXene spin density while the "down" polarisation interacts with the nearest titanium atoms and flips their polarisation. Therefore, the spin density around the upper layer in Ti\textsubscript{2}C drops. 

\begin{figure}
  \centering
  \includegraphics[width=\columnwidth]{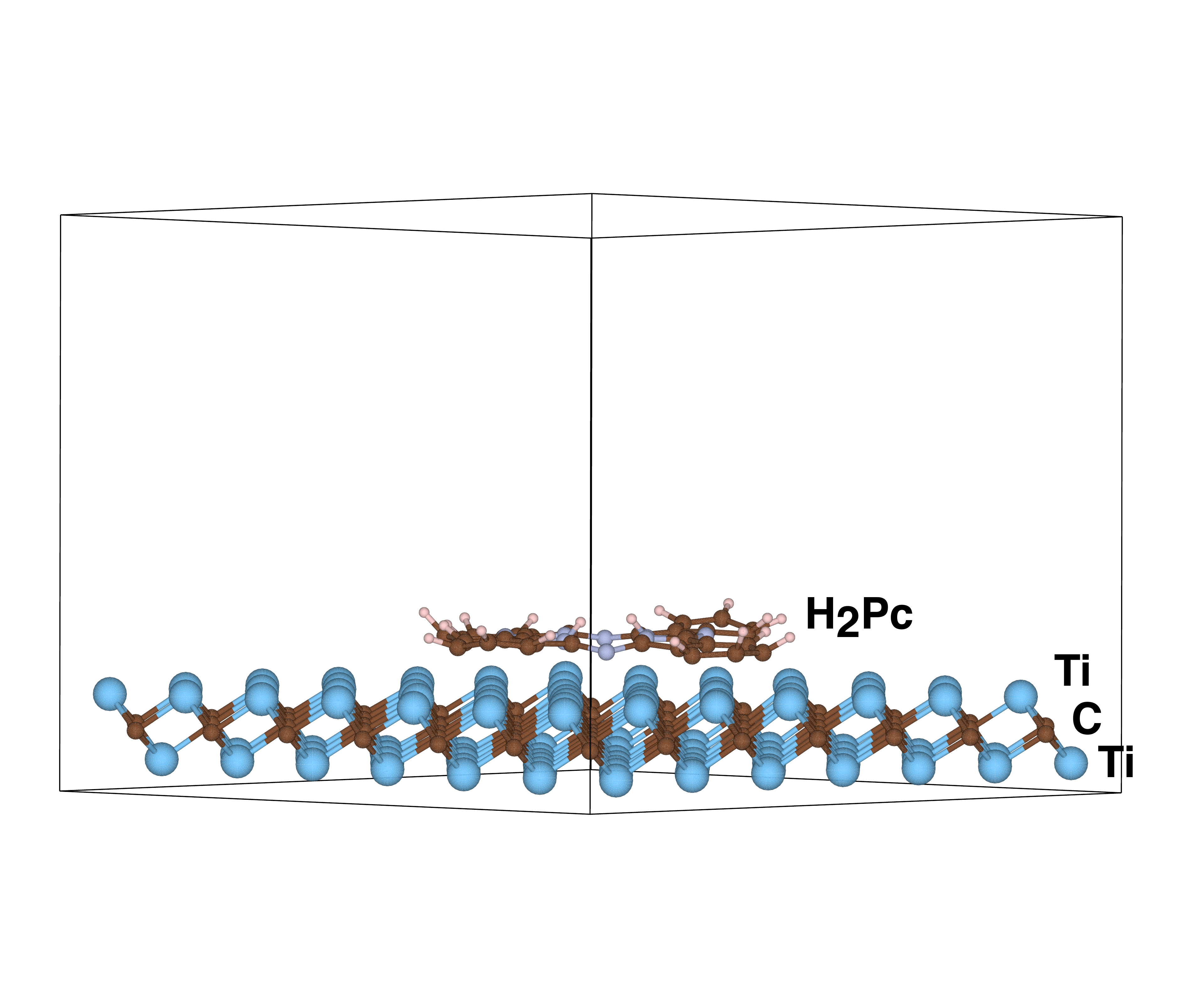}
  \caption{Side view of the H\textsubscript{2}Pc/Ti\textsubscript{2}C hybrid system.}
  \label{fig:H2Ti2CView}
\end{figure}

To understand the source of the high adsorption energy of FePc on Ti\textsubscript{2}C, the H\textsubscript{2}Pc molecule was optimised on the same layer. Fig. \ref{fig:H2Ti2CView} shows that two hydrogen atoms that substituted the iron atom are pushed off from the layer during the optimisation. The comparison of the geometric and energetic parameters (Table \ref{table:H2Ti2CCharact}) shows that the adsorption energies of the H\textsubscript{2}Pc and FePc on the layer are similar and the adsorption energy is not additive. The adsorption is mostly caused by adhesion of the Pc ligand to the Ti metallic layer rather than by adhesion of the iron atom.

\begin{table}[h!t]
\caption{Comparison of energetic and geometric characteristics of H\textsubscript{2}Pc, Fe and FePc on Ti\textsubscript{2}C layer}
\label{table:H2Ti2CCharact}
        \centering
\begin{ruledtabular}
\begin{tabular}{cccc}
  & $E_a$, eV \ \  & \makecell{Pc - Ti\textsubscript{2}C\\dist, \AA} & \makecell{Fe - Ti\textsubscript{2}C\\dist, \AA} \\
\colrule
 H\textsubscript{2}Pc/Ti\textsubscript{2}C AF & -21.75 & 2.02 & --- \\
 Fe/Ti\textsubscript{2}C & -4.02 & --- & \makecell{1,83 FeUp\\1,77 FeDown} \\
 FePc/Ti\textsubscript{2}C & -21.88 & 2.09 & 2.03 \\
\end{tabular}
\end{ruledtabular}
        \end{table}

\section{Conclusions}

The early stage study of a TMPc molecule on a magnetic MXene layer has been made. This investigation of the FePc/Ti\textsubscript{2}C hybrid system can serve as a starting point for other similar studies of organometallic molecules on MXene layers. 
    
FePc adsorption to Ti\textsubscript{2}C is much stronger compared to other 2D materials, such as MoS\textsubscript{2} and graphene.\cite{haldar2018comparative} Adsorption is no longer determined by the physical van der Waals interaction, but by chemical bonds between the carbon rings and the titanium surface. The iron atom makes a small contribution to adsorption. 

The exchange energy $E_ex$, i.e. the energy difference between the cases with the same initial Ti\textsubscript{2}C magnetisation and different orientation of the iron atom in the FePc molecule, is 283 meV for the ferromagnetic case, 9 meV for the antiferromagnetic case, and 0.6 meV for the antiferromagnetic case with the molecule in the excited state.
For the iron atom on the antiferromagnetic Ti\textsubscript{2}C layer with the same supercell size, it is 430 meV. The special thing about FePc is that by changing the spin orientation, the molecular geometry does not change, while in Fe/Ti\textsubscript{2}C the flip of the spin orientation of the iron atom is associated with the movement of the iron atom along the $z$-axis.

The ferromagnetic interaction between the iron atom (and the iron atom in FePc) and the upper Ti layer in MXene was found. When spin polarisations of the upper Ti layer in Ti\textsubscript{2}C  and the Fe atom are co-directed, titanium atoms closest to the iron atom have a tendency to flip their spin polarisation.

Big charge transfer (about 8.5 $\bar{e}$) from Ti\textsubscript{2}C to FePc is observed in the hybrid system. Most of this charge is transferred to FePc carbon atoms while about 0.38 electron is transferred to the iron atom.  For comparison, charge transfer from the gold layer to the VPc molecule is just 0.62 $\bar{e}$.\cite{mabrouk2021stability}

\section{Acknowledgement}
This work has been supported by National Science Centre Poland (UMO 2016/23/B/ST3/03567).

\bibliography{Article}

\end{document}


\preprint{AIP/123-QED}

\title{Supplemental material}%



\maketitle

\subsection{Density of states analysis}

\begin{figure}[!ht]
\centering
\begin{subfigure}{0.49\linewidth}
  \centering
  \includegraphics[width=\linewidth]{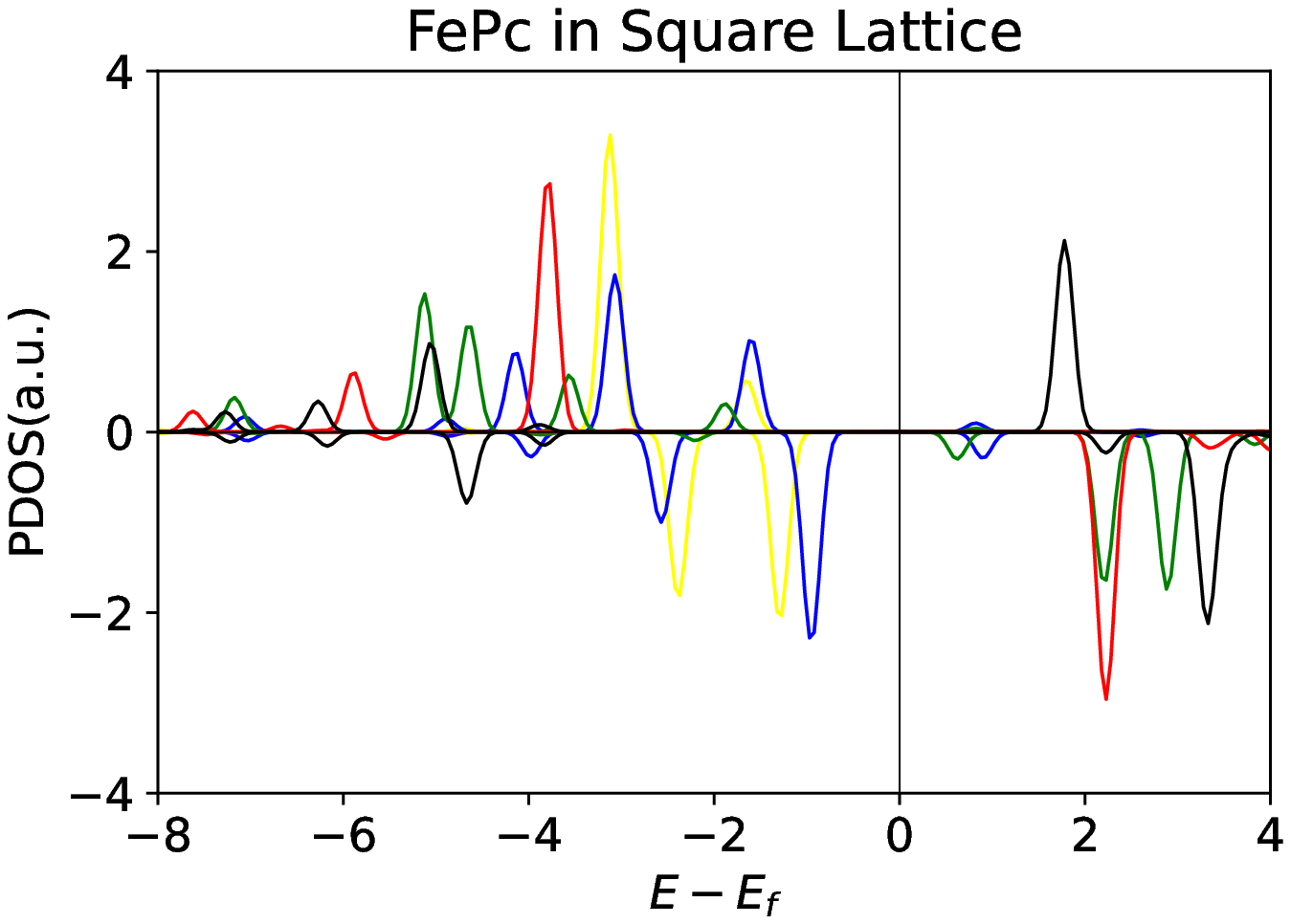}
  \caption{}
  \label{fig:FePcPDOS}
\end{subfigure}
\begin{subfigure}{0.49\linewidth}
  \centering
  \includegraphics[width=\linewidth]{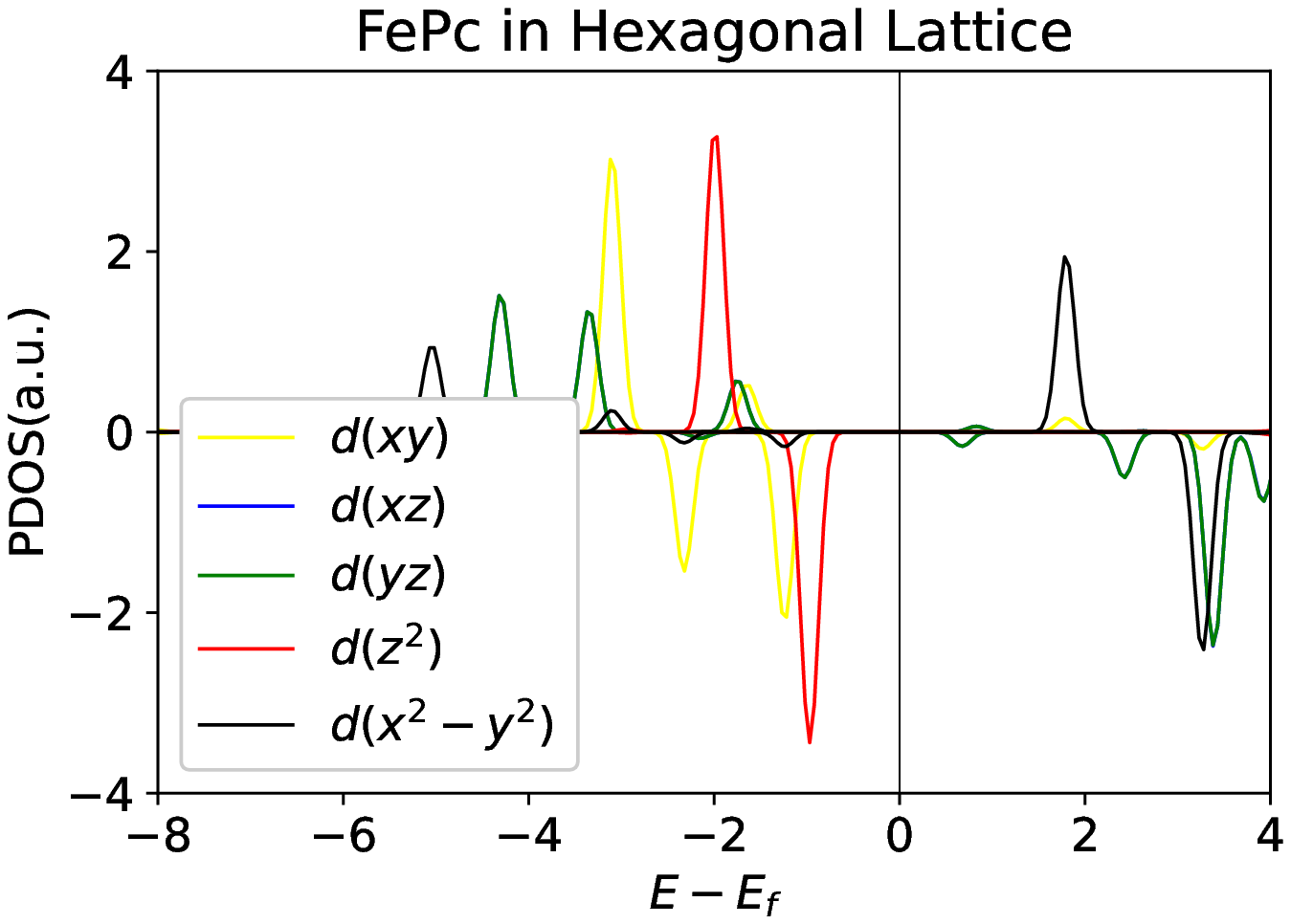}
  \caption{}
  \label{fig:FePcPDOSHex}
\end{subfigure}
\begin{subfigure}{0.4\linewidth}
\centering
\includegraphics[width=\linewidth]{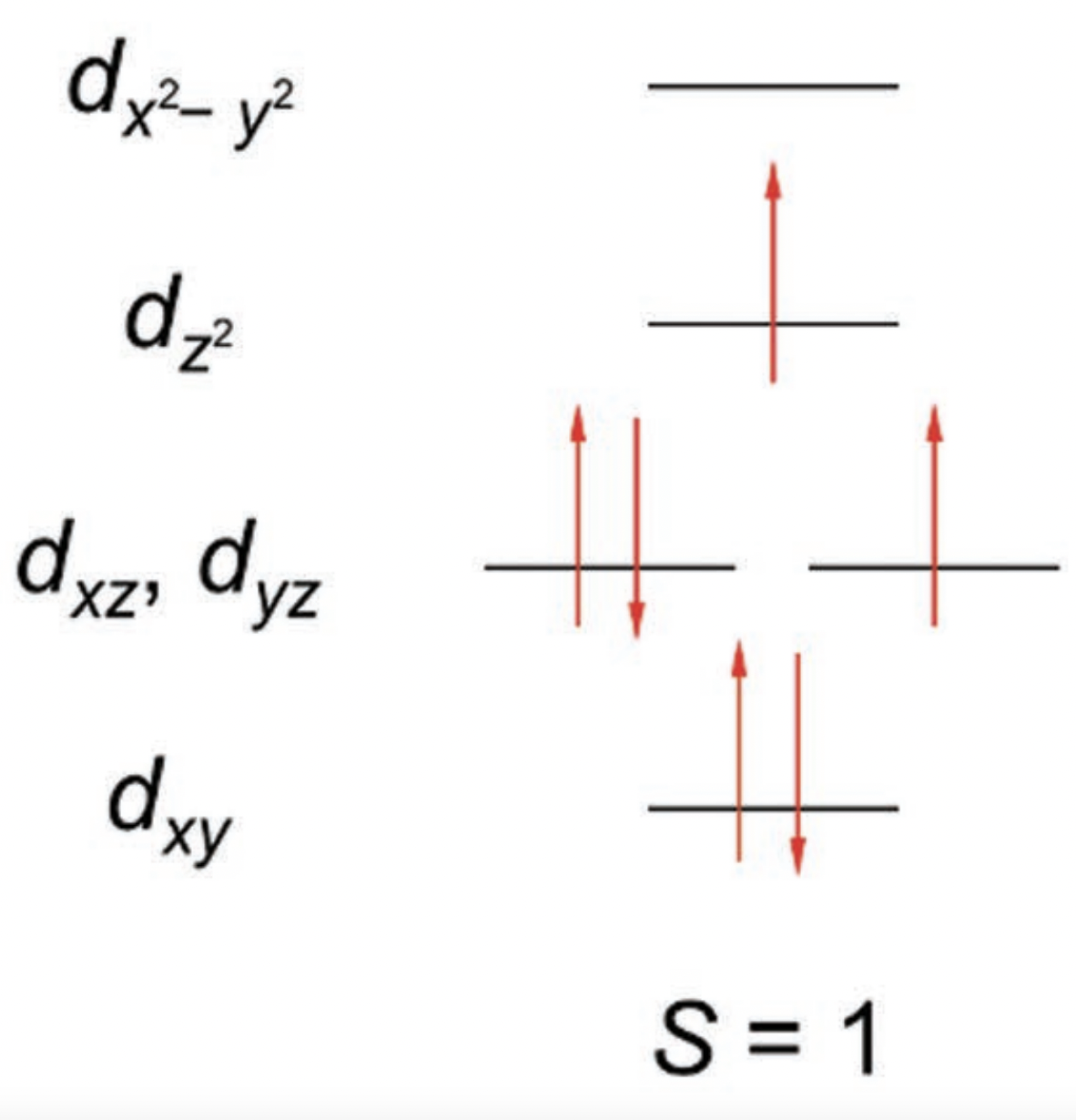}
\caption{}
\label{fig:FePcdOrb}
\end{subfigure}
\begin{subfigure}{0.4\linewidth}
  \centering
  \includegraphics[width=\linewidth]{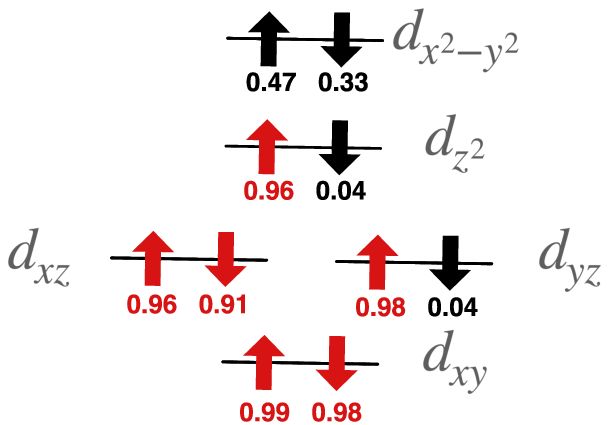}
  \caption{}
  \label{fig:FePcDorb}
\end{subfigure}

\caption{Analysis of the iron $d$-orbitals in FePc. Spin resolved projected onto $d$-states density of states for FePc molecule in (a) cubic and (b) hexagonal cells; (c) iron $3d$ shell configuration in the $E_g$ FePc ground state; (d) Scheme of energetically split $d$-orbitals in the crystal field of tetragonal symmetry. The Löwdin charges associated with each spin up and down $d$-orbital are given below the levels. The six orbitals with the highest charge (i.e., occupation probability) are indicated in red colour.
}
\label{fig:FePcPDOSDorb}
\end{figure}

\begin{figure}[!ht]
\centering
\begin{subfigure}{0.49\linewidth}
  \centering
  \includegraphics[width=\linewidth]{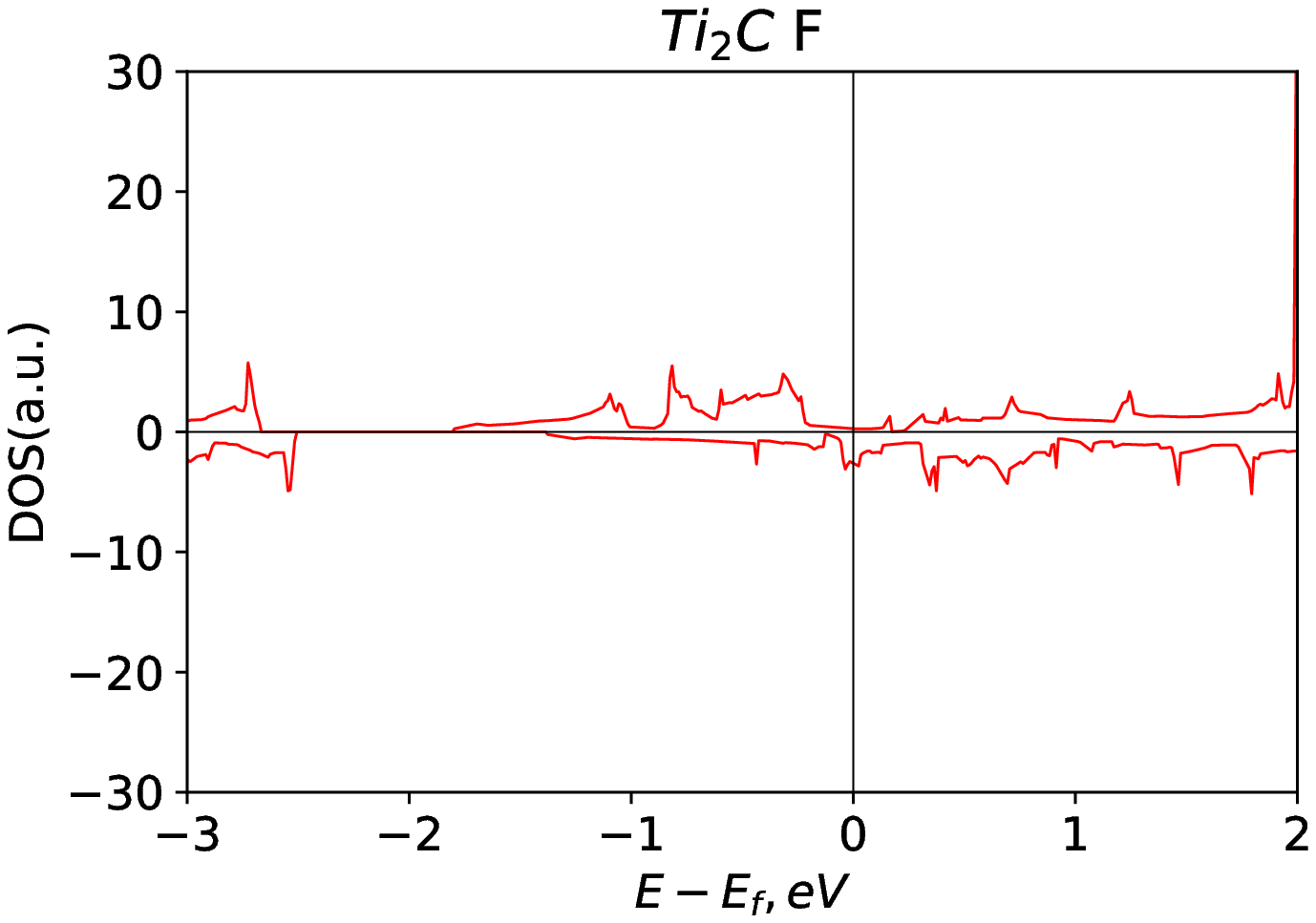}
  \caption{}
  \label{fig:Ti2CFDOS}
\end{subfigure}
\hfill
\begin{subfigure}{0.49\linewidth}
  \centering
  \includegraphics[width=\linewidth]{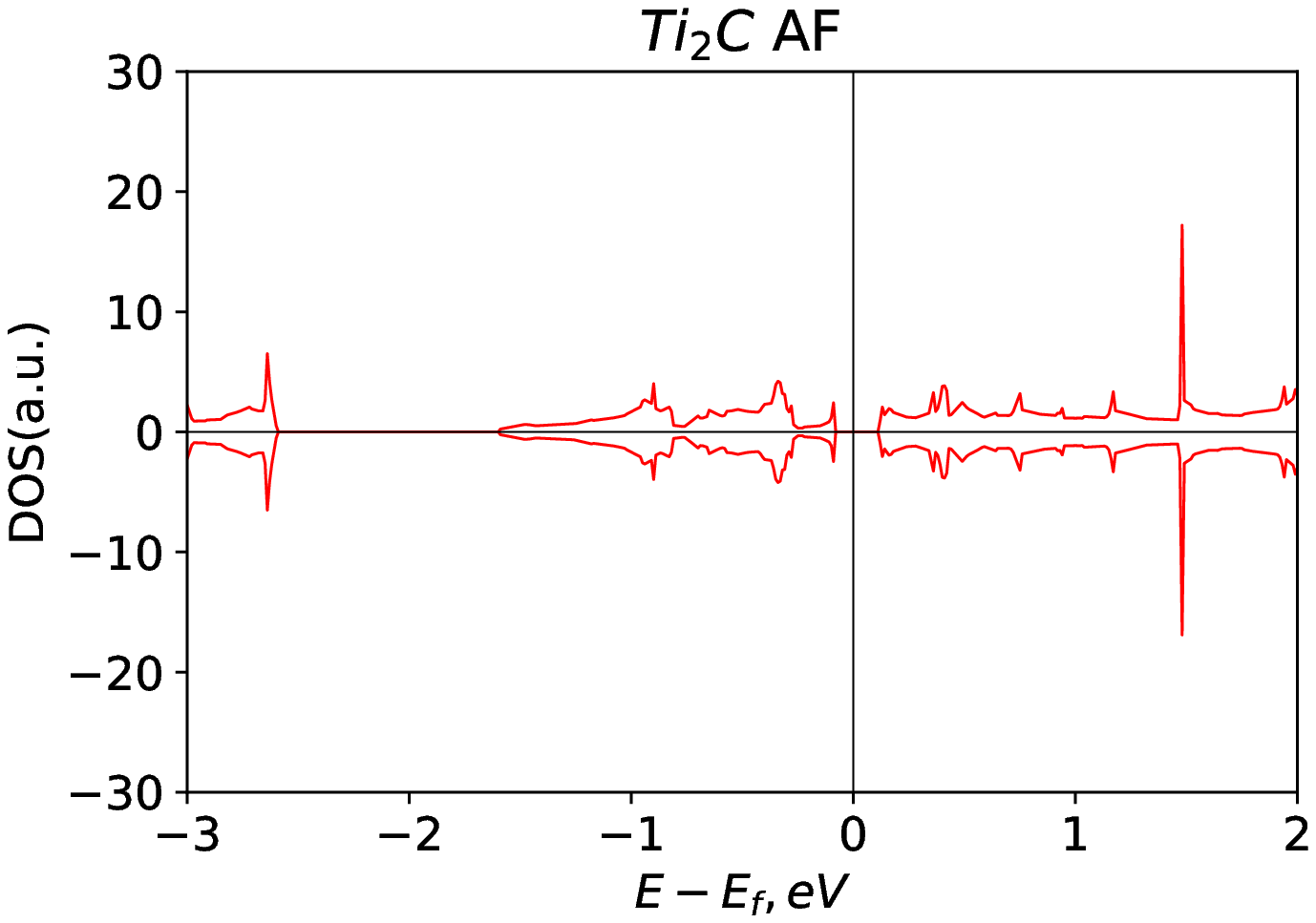}
  \caption{}
  \label{fig:Ti2CAFDOS}
\end{subfigure}
\caption{Spin up \& down densities of states for  Ti\textsubscript{2}C trilayer with (a) ferromagnetic, and (b) antiferromagnetic ordering of magnetic moments in the Ti sublayers. Note small band gap in Ti\textsubscript{2}C with the antiferromagnetic ordering of Ti layers.}
\label{fig:Ti2CDOS}
\end{figure}

Before we discuss the density of states in the FePc/Ti\textsubscript{2}C hybrid system, let us have a look at the $d$-level splitting and the PDOS of the free standing FePc depicted in Fig. \ref{fig:FePcPDOSDorb} and the DOS of Ti\textsubscript{2}C with ferromagnetic and antiferromagnetic orderings of Ti layers as depicted in Fig. \ref{fig:Ti2CDOS}.






In the DFT calculations for free standing FePc we employed square based supercell, which guarantees that the symmetry is really tetragonal. Since the symmetry of free standing FePc is tetragonal, and the symmetry of Ti\textsubscript{2}C is trigonal, the symmetry of the hybrid system is very low, namely monoclinic. Therefore, the iron $d_{xz}$ and $d_{yz}$ orbitals, which are energetically degenerated in the free standing FePc molecule, will split in the FePc/Ti\textsubscript{2}C hybrid system.

In Fig. \ref{fig:FePcPDOSDorb} it is clearly seen that the sum of both spin occupancies over the six most popular populated levels gives spin value S=1, whereas this value in the free Fe atom is S=2.

Concerning the electronic structure of Ti\textsubscript{2}C, we point out that the structure with antiferromagnetic ordering of magnetic moments in Ti layers exhibits tiny band gap. 

\begin{figure}[ht!]
\centering
\begin{subfigure}{0.49\linewidth}
  \centering
  \includegraphics[width=\linewidth]{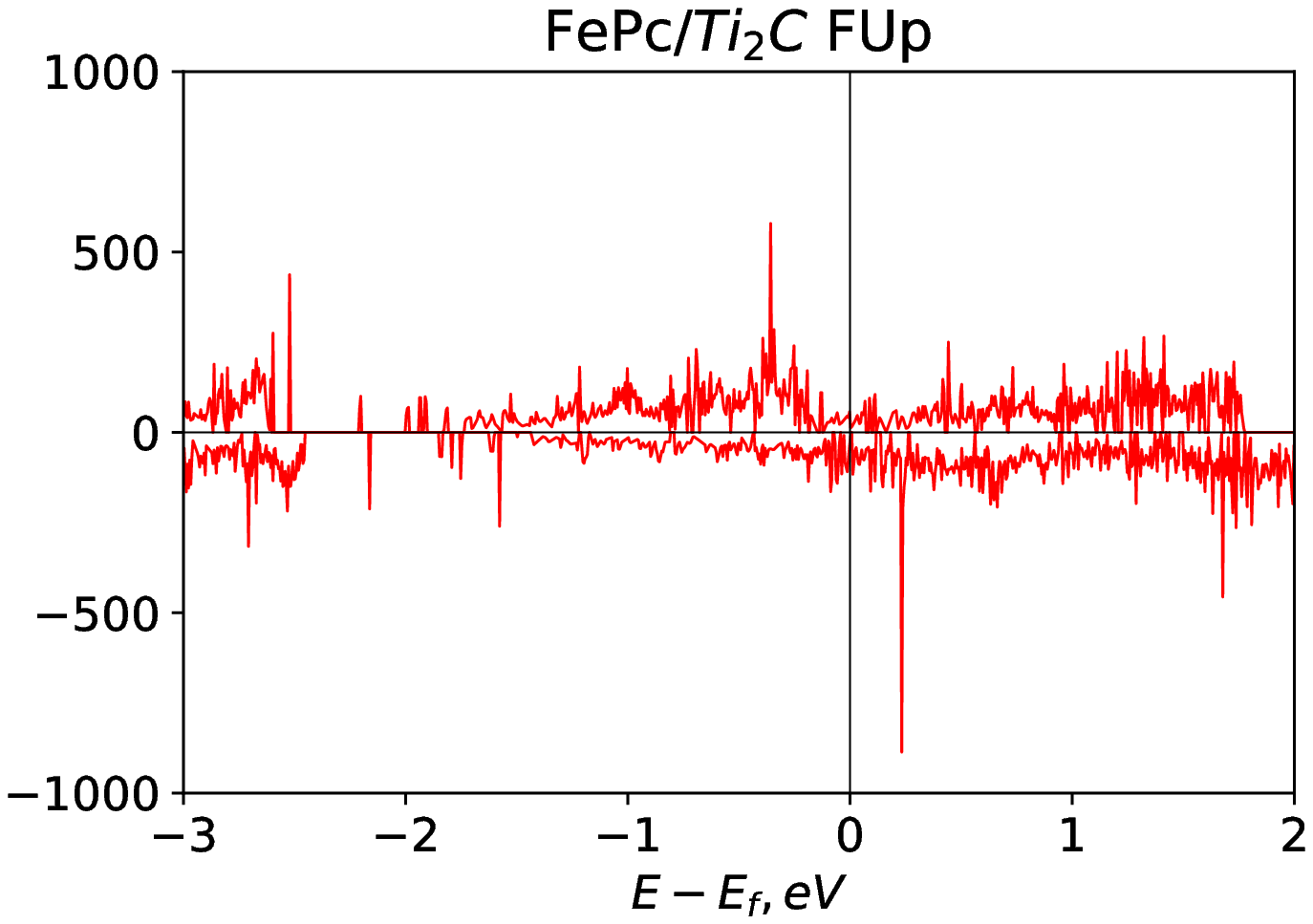}
  \caption{}
  \label{fig:FUpU4DOS}
\end{subfigure}
\hfill
\begin{subfigure}{0.49\linewidth}
  \centering
  \includegraphics[width=\linewidth]{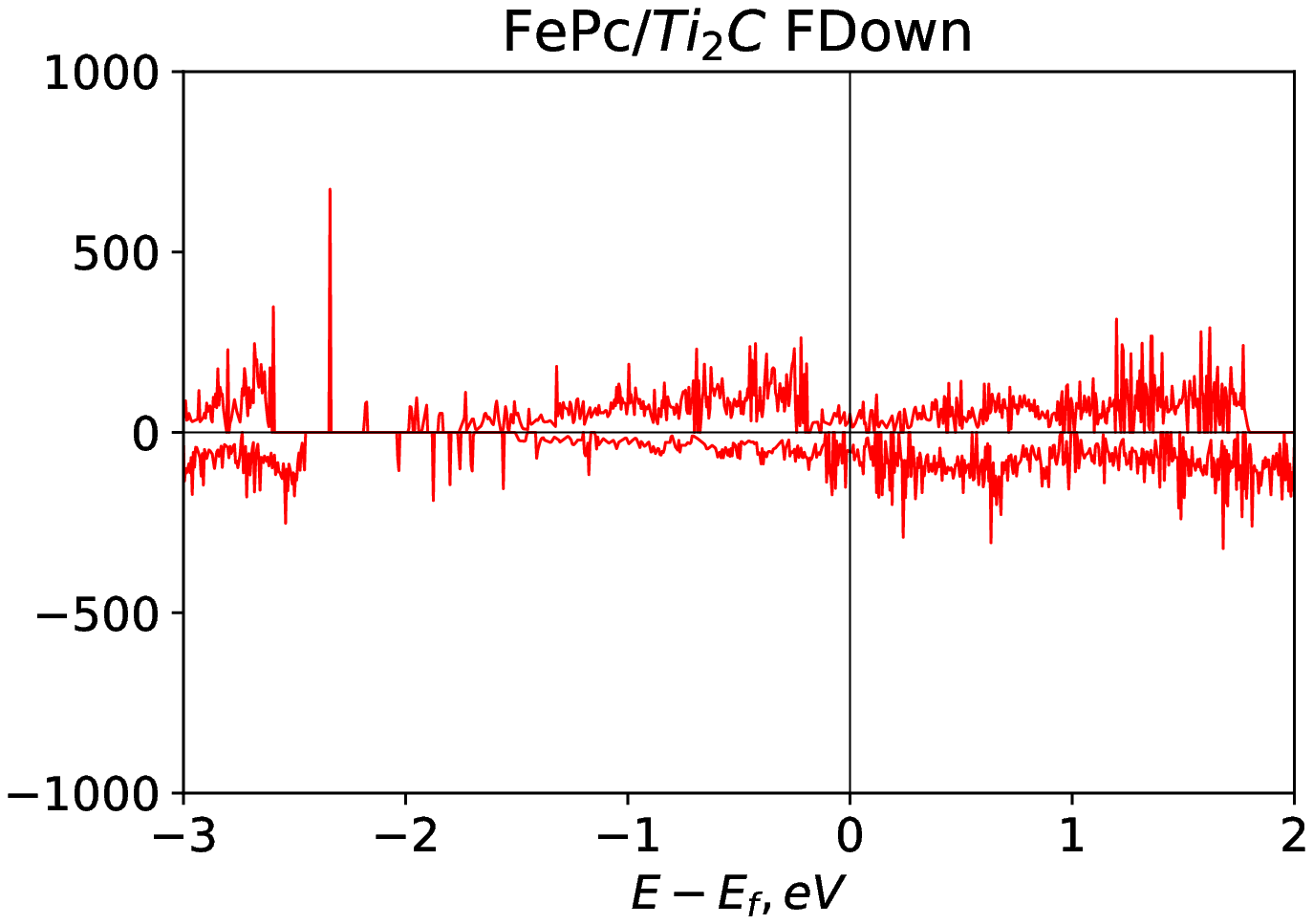}
  \caption{}
  \label{fig:FDownU4DOS}
\end{subfigure}

\begin{subfigure}{0.49\linewidth}
  \centering
  \includegraphics[width=\linewidth]{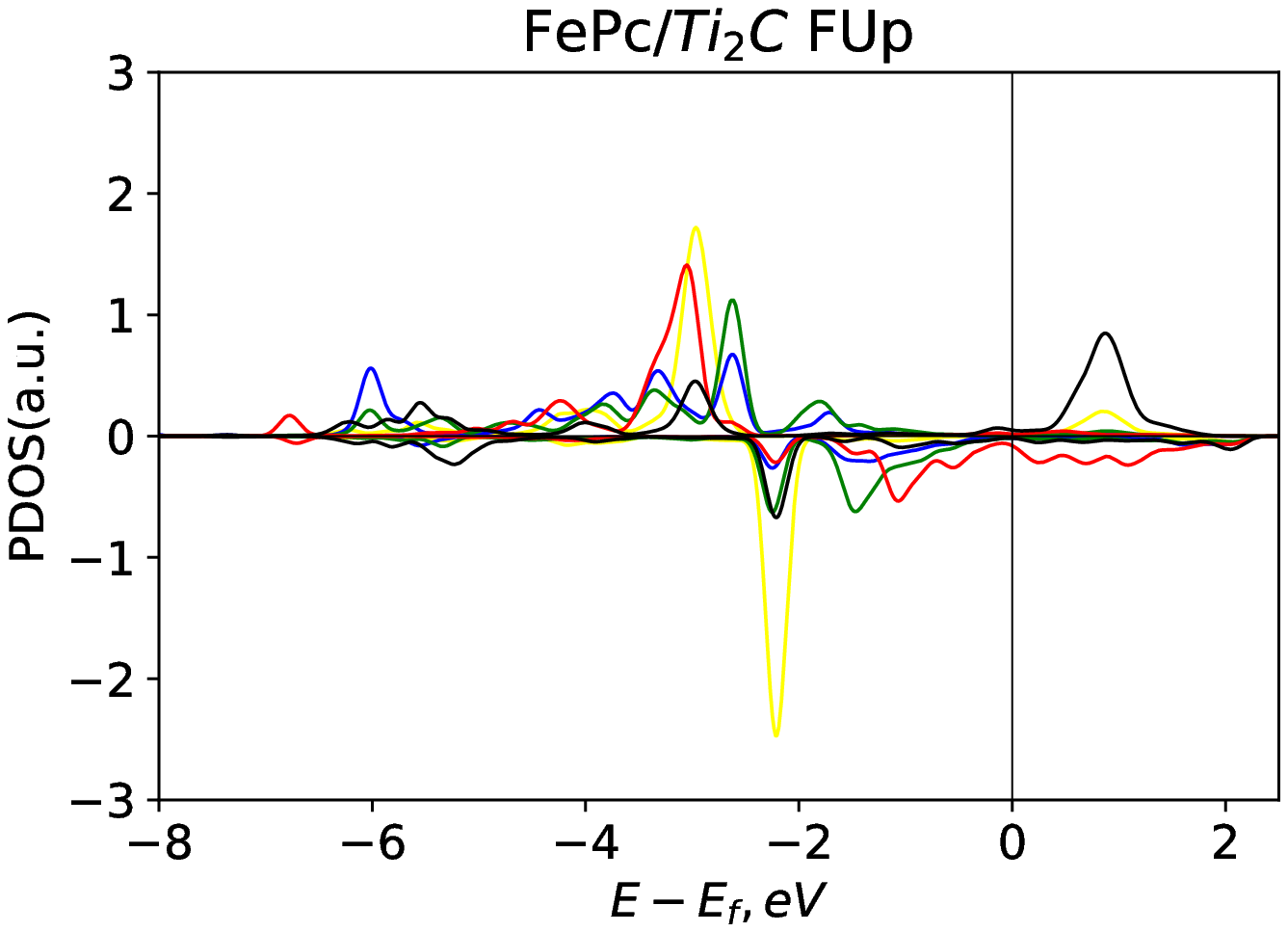}
  \caption{}
  \label{fig:FUpU4PDOS}
\end{subfigure}
\hfill
\begin{subfigure}{0.49\linewidth}
  \centering
  \includegraphics[width=\linewidth]{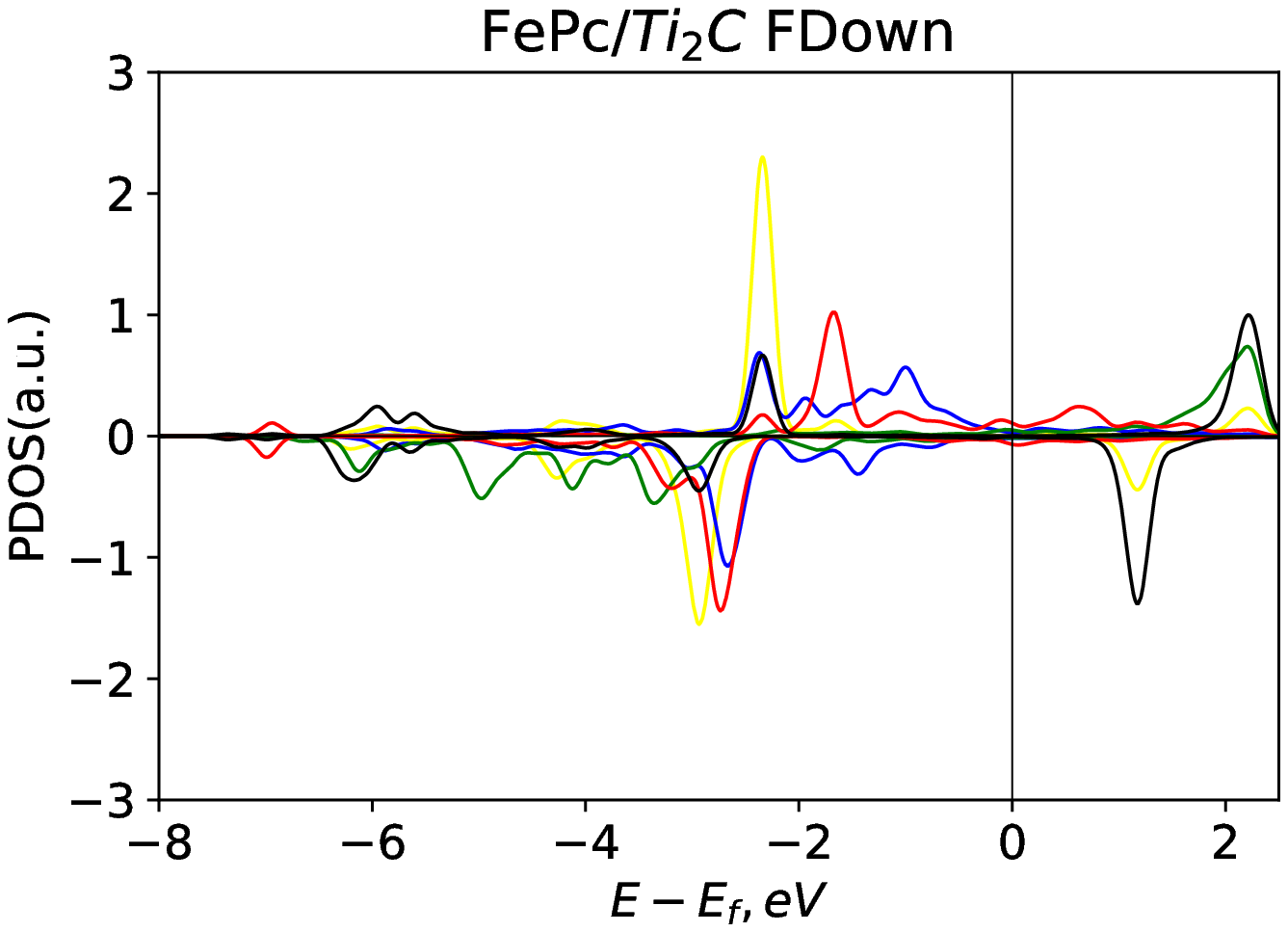}
  \caption{}
  \label{fig:FDownU4PDOS}
\end{subfigure}

\begin{subfigure}{0.4\linewidth}
  \centering
  \includegraphics[width=\textwidth]{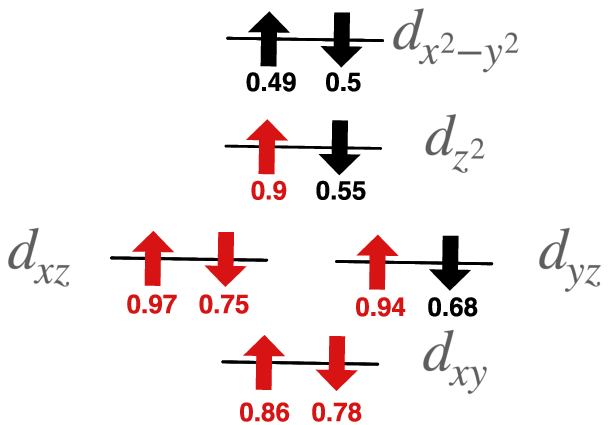}
  \caption{}
  \label{fig:FUpU4DOrb}
\end{subfigure}
\hfill
\begin{subfigure}{0.4\linewidth}
  \centering
  \includegraphics[width=\textwidth]{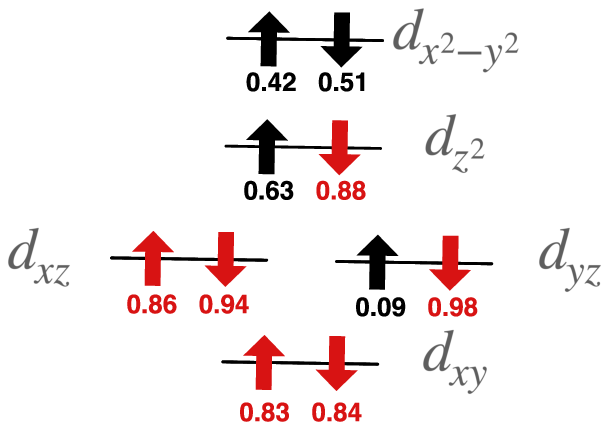}
  \caption{}
  \label{fig:FDownU4DOrb}
\end{subfigure}

\caption{The electronic structure analysis of the FePc/Ti\textsubscript{2}C systems \textit{with the Ti\textsubscript{2}C ferromagnetic configuration} FUp and FDown: total density of states for (a) iron spin up and (b) iron spin down cases, the iron d-orbital projected densities of states for (c) iron spin up and (d) iron spin down cases (yellow - ${d_{xy}}$, blue - ${d_{xz}}$, green - $d_{yz}$, red - $d_{z^2}$, black - $d_{x^2-y^2}$); Löwdin charges of each d-orbital component in the iron atom for (e) iron spin up and (f) iron spin down cases. The six orbitals with the highest occupation numbers are marked in red. For clarity of the PDOS pictures, the gaussian broadening parameter was taken to be equal to the energy grid step (0.005 eV).}
\label{fig:FerromU4}
\end{figure}

\begin{figure}[ht!]
\centering
\begin{subfigure}{0.49\linewidth}
  \centering
  \includegraphics[width=\linewidth]{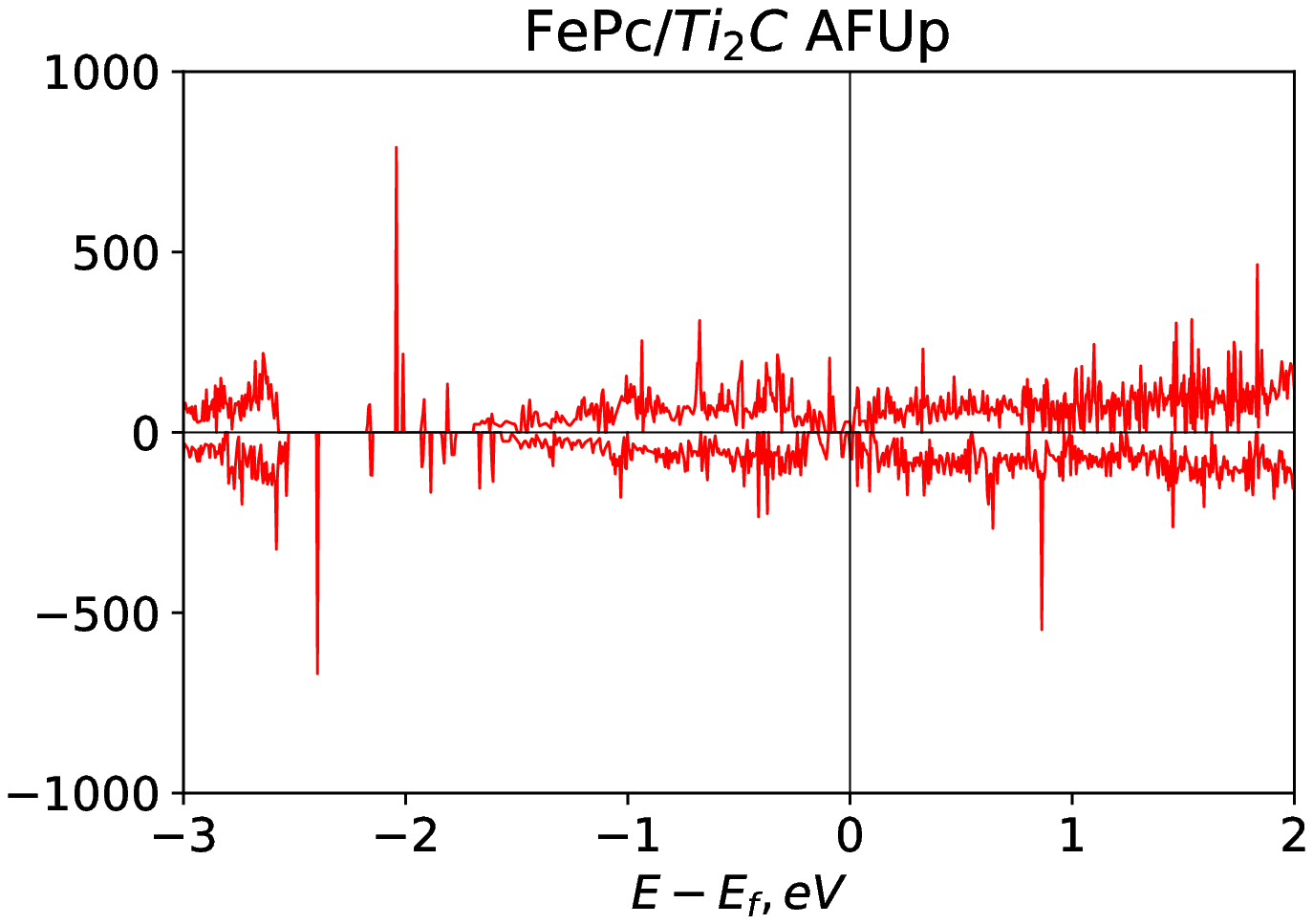}
  \caption{}
  \label{fig:AFUpU4DOS}
\end{subfigure}
\hfill
\begin{subfigure}{0.49\linewidth}
  \centering
  \includegraphics[width=\linewidth]{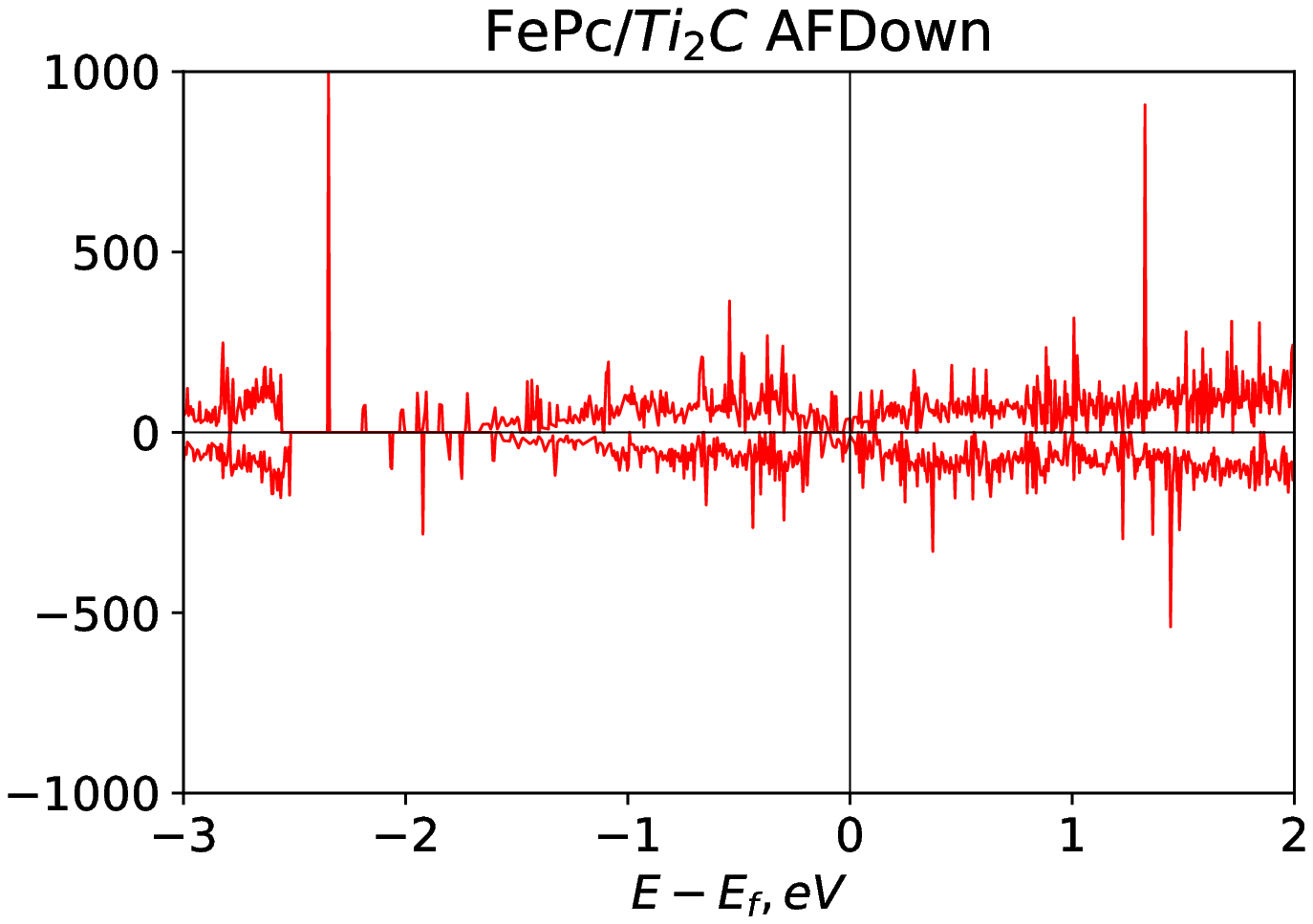}
  \caption{}
  \label{fig:AFDownU4DOS}
\end{subfigure}

\begin{subfigure}{0.49\linewidth}
  \centering
  \includegraphics[width=\linewidth]{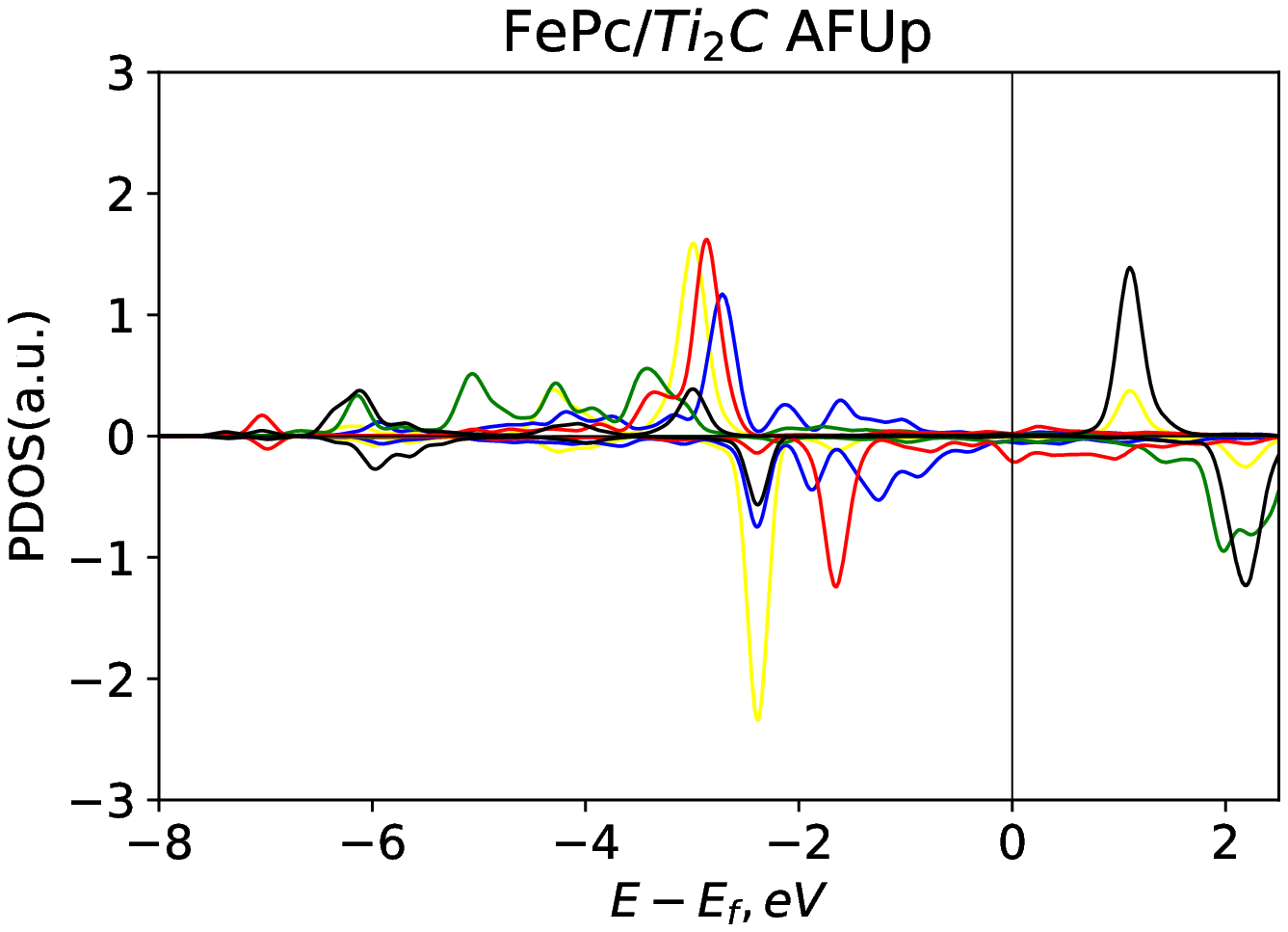}
  \caption{}
  \label{fig:AFUpU4PDOS}
\end{subfigure}
\hfill
\begin{subfigure}{0.49\linewidth}
  \centering
  \includegraphics[width=\linewidth]{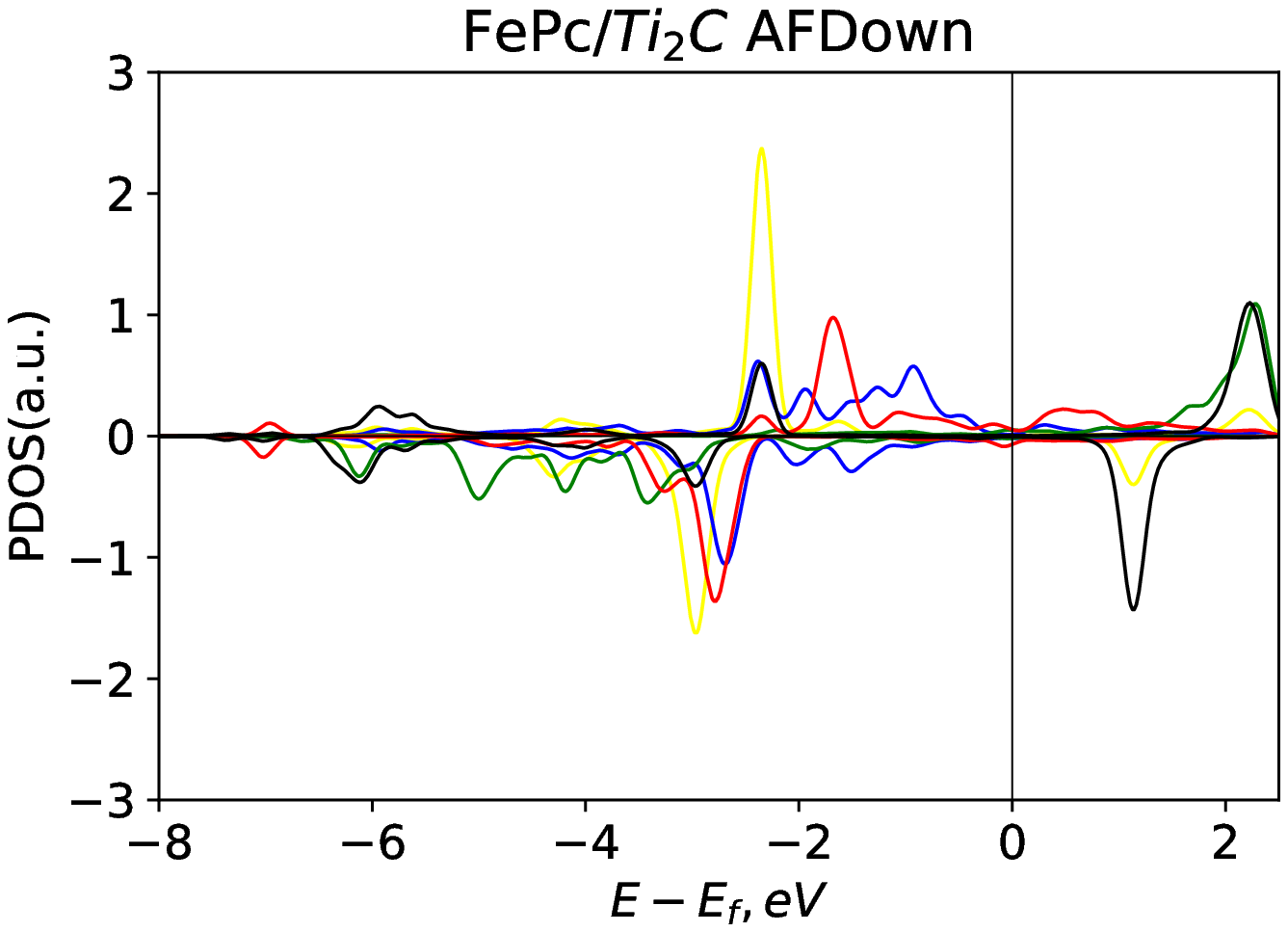}
  \caption{}
  \label{fig:AFDownU4PDOS}
\end{subfigure}

\begin{subfigure}{0.4\linewidth}
  \centering
  \includegraphics[width=\textwidth]{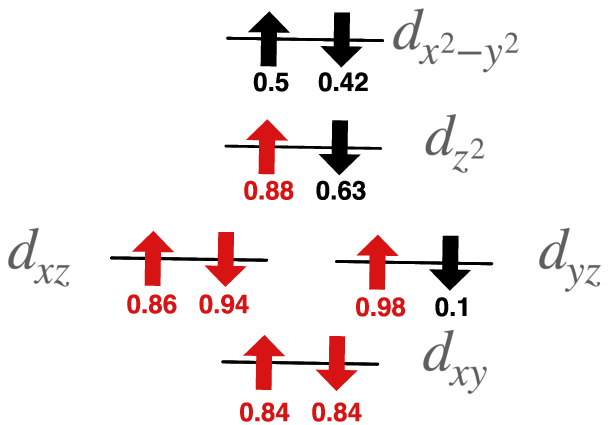}
  \caption{}
  \label{fig:AFUpU4DOrb}
\end{subfigure}
\hfill
\begin{subfigure}{0.4\linewidth}
  \centering
  \includegraphics[width=\textwidth]{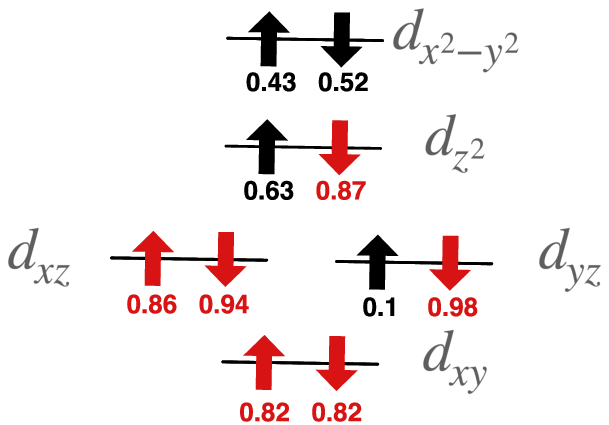}
  \caption{}
  \label{fig:AFDownU4DOrb}
\end{subfigure}

\caption{The electronic structure analysis of the FePc/Ti\textsubscript{2}C systems \textit{with the Ti\textsubscript{2}C antiferromagnetic configuration} AFUp and AFDown: total density of states for (a) iron spin up and (b) iron spin down cases, the iron d-orbital projected densities of states for (c) iron spin up and (d) iron spin down cases (yellow - ${d_{xy}}$, blue - ${d_{xz}}$, green - $d_{yz}$, red - $d_{z^2}$, black - $d_{x^2-y^2}$); Löwdin charges of each d-orbital component in the iron atom for (e) iron spin up and (f) iron spin down cases. The six orbitals with the highest occupation numbers are marked in red. For clarity of the PDOS pictures, the gaussian broadening parameter was taken to be equal to the energy grid step (0.005 eV). }
\label{fig:AntiFerromU4}
\end{figure}

\begin{figure}[ht!]
\centering
\begin{subfigure}{0.49\linewidth}
  \centering
  \includegraphics[width=\linewidth]{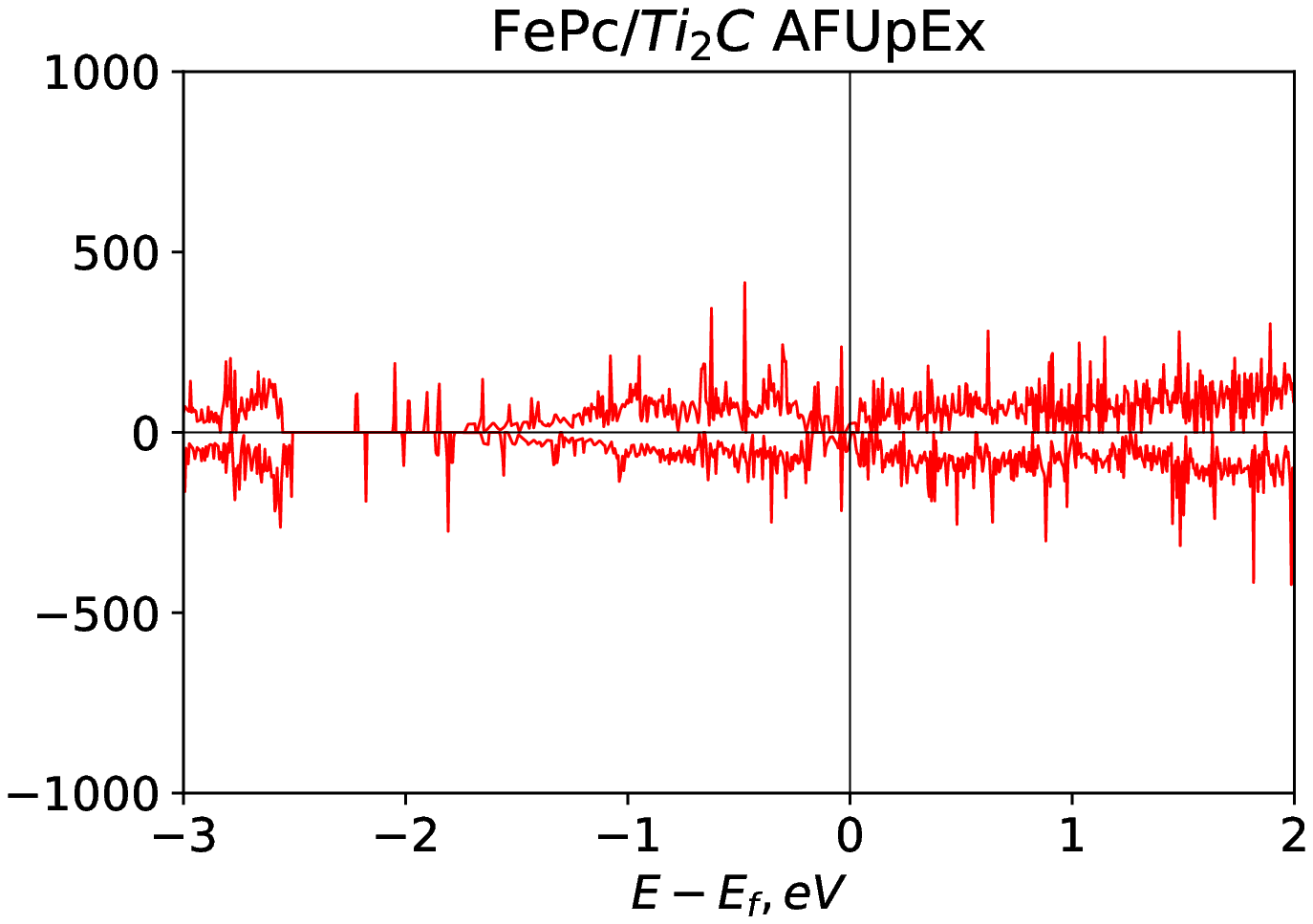}
  \caption{}
  \label{fig:AFUpU6DOS}
\end{subfigure}
\hfill
\begin{subfigure}{0.49\linewidth}
  \centering
  \includegraphics[width=\linewidth]{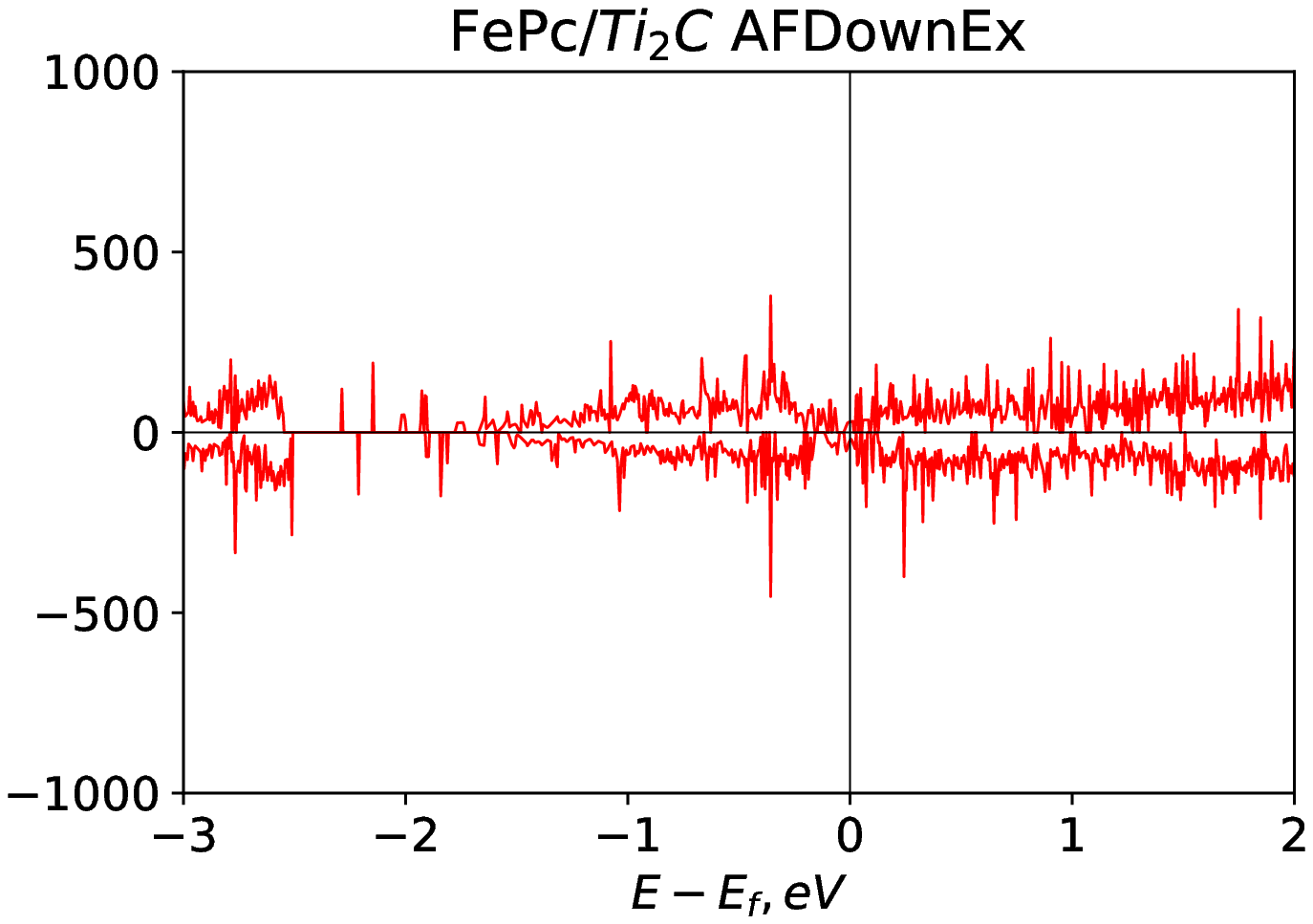}
  \caption{}
  \label{fig:AFDownU6DOS}
\end{subfigure}

\begin{subfigure}{0.49\linewidth}
  \centering
  \includegraphics[width=\linewidth]{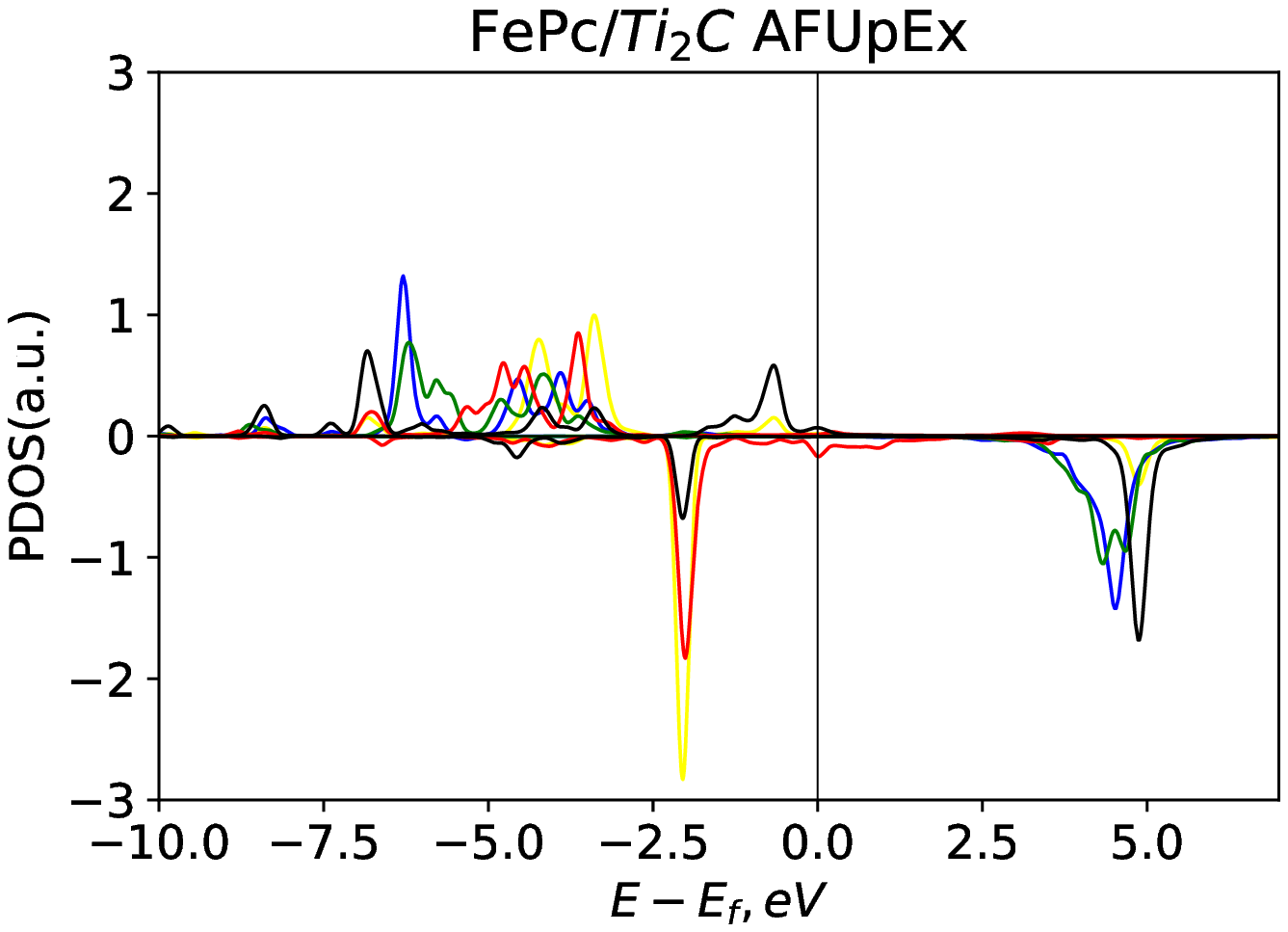}
  \caption{}
  \label{fig:AFUpU6PDOS}
\end{subfigure}
\hfill
\begin{subfigure}{0.49\linewidth}
  \centering
  \includegraphics[width=\linewidth]{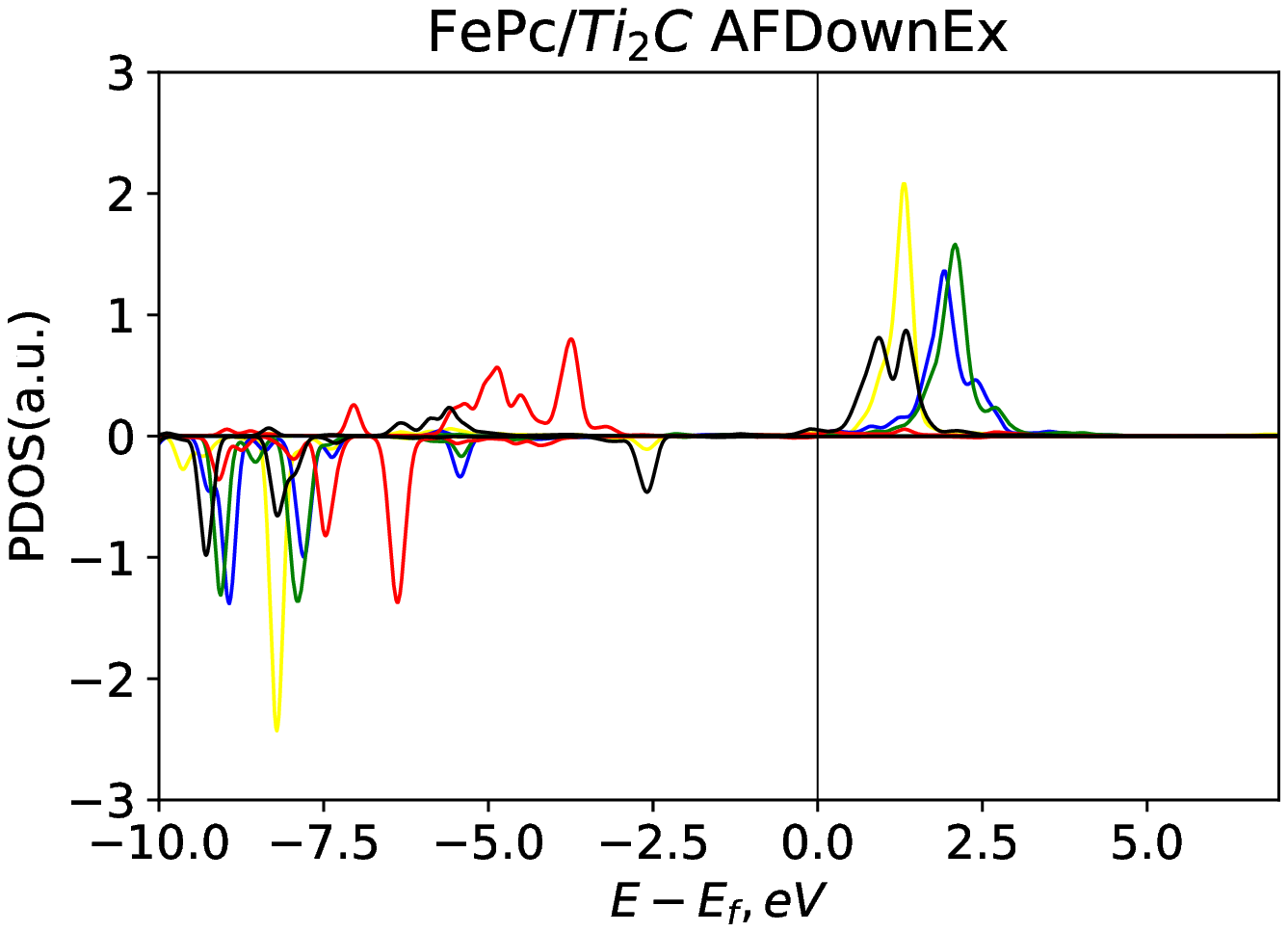}
  \caption{}
  \label{fig:AFDownU6PDOS}
\end{subfigure}

\begin{subfigure}{0.4\linewidth}
  \centering
  \includegraphics[width=\textwidth]{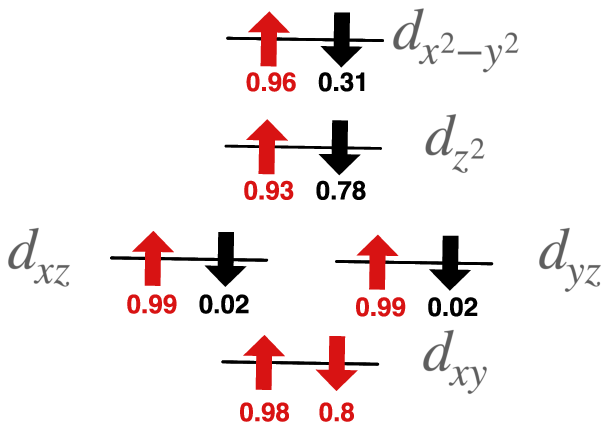}
  \caption{}
  \label{fig:AFUpU6DOrb}
\end{subfigure}
\hfill
\begin{subfigure}{0.4\linewidth}
  \centering
  \includegraphics[width=\textwidth]{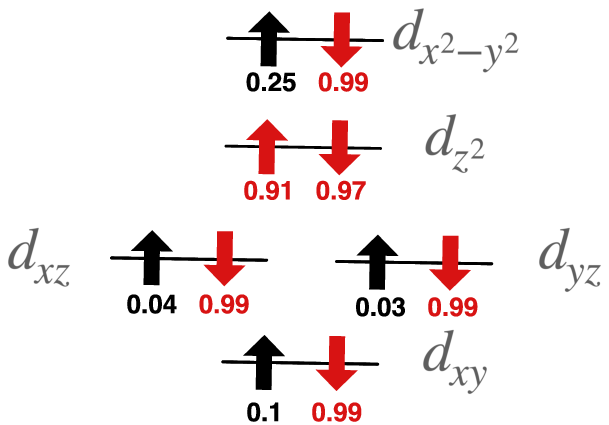}
  \caption{}
  \label{fig:AFDownU6DOrb}
\end{subfigure}

\caption{The electronic structure analysis of the FePc/Ti\textsubscript{2}C systems \textit{with the Ti\textsubscript{2}C antiferromagnetic configuration and FePc in the excited state} AFUpEx and AFDownEx: total density of states for (a) iron spin up and (b) iron spin down cases, the iron d-orbital projected densities of states for (c) iron spin up and (d) iron spin down cases (yellow - ${d_{xy}}$, blue - ${d_{xz}}$, green - $d_{yz}$, red - $d_{z^2}$, black - $d_{x^2-y^2}$); Löwdin charges of each d-orbital component in the iron atom for (e) iron spin up and (f) iron spin down cases. The six orbitals with the highest occupation numbers are marked in red. For clarity of the PDOS pictures, the gaussian broadening parameter was taken to be equal to the energy grid step (0.005 eV).}
\label{fig:AntiFerromU6}
\end{figure}

Now, we are in the position to perform an analysis of the DOS for all considered variants of the FePc/Ti\textsubscript{2}C hybrid systems named as FUp, FDown, AFUp, and AFDown. The results are presented in the series of figures: Fig. \ref{fig:FerromU4} for FUp and FDown hybrid systems; Fig. \ref{fig:AntiFerromU4} for the AFUp and AFDown hybrid systems; Fig. \ref{fig:AntiFerromU6} for the AFUp and AFDown hybrid systems with the FePc molecule in the excited state indicated as AFUpEx and AFDownEx.

At the first glance on Figs. \ref{fig:FerromU4}, \ref{fig:AntiFerromU4}, and \ref{fig:AntiFerromU6} one realises that all investigated FePc/Ti\textsubscript{2}C hybrid structures are metallic in contrast to the Ti\textsubscript{2}C with antiferromagnetic ordering of Ti layers that exhibits a small band gap.

At a closer look at projected densities of states for the studied hybrid systems (see Figs. \ref{fig:FerromU4}, \ref{fig:AntiFerromU4}, and \ref{fig:AntiFerromU6}, panels (C) and (D)), one realises that the electronic energy levels originating from the iron $d$-orbitals are placed fairly close to the Fermi energy. 
The population analysis of the iron d-orbitals in the FUp, FDown, AFUp, and AFDown forms of the FePc/Ti\textsubscript{2}C hybrid systems is schematically shown in Fig. \ref{fig:FUpU4DOrb}, \ref{fig:FDownU4DOrb}, \ref{fig:AFUpU4DOrb}, \ref{fig:AFDownU4DOrb}. The $d$-orbital populations of those systems with FePc in its ground state qualitatively coincide with the populations for the isolated FePc molecule (Fig. \ref{fig:FePcPDOSDorb}). For the molecule in the excited state (Fig. \ref{fig:AFUpU6DOrb}, \ref{fig:AFDownU6DOrb}), i.e., for hybrid systems indicated as AFUpEx and AFDownEx, it can be seen that the upper orbital $d_{x^2-y^2}$ is populated by one electron. Also, in both ground and excited states, the $d_{z^2}$ orbital is completely populated. The iron atom leaves the plane of the molecule, the ligand field changes, and the electronegative nitrogen atoms are no longer in the $xy$-plane. As a consequence, the $d_{z^2}$ orbital becomes more energetically favourable.

\clearpage

\clearpage

\clearpage

\clearpage
